\documentclass[12pt,epsf,amssymb,citesort]{article}
\usepackage{tabularx}
\usepackage{array}
\usepackage{graphics}
\usepackage{graphicx}
\usepackage{epsfig}
\usepackage{amsmath}
\usepackage{amssymb}
\usepackage{axodraw}
\usepackage{cite}
\makeatletter

\usepackage{verbatim}

\newcommand{\bea}{\begin{eqnarray}}
\newcommand{\eea}{\end{eqnarray}}
\def\beq{\begin{equation}}
\def\eeq{\end{equation}}

\newcommand{\pdir}{p\kern -5.2pt\raise 0.2ex\hbox {/}}
\newcommand{\vdir}{v\kern -5.75pt\raise 0.15ex\hbox {/}}
\newcommand{\kdir}{k\kern -5.75pt\raise 0.15ex\hbox {/}}
\newcommand{\epsdir}{\epsilon\kern -5.0pt\raise 0.15ex\hbox {/}}
\newcommand{\bvdir}{\bar{v}\kern -5.75pt\raise 0.15ex\hbox {/}}
\newcommand{\Ddir}{D\kern -7.75pt\raise 0.20ex\hbox {/}}
\newcommand{\Adir}{A\kern -7.75pt\raise 0.20ex\hbox {/}}
\newcommand{\ldir}{l\kern -5.0pt\raise 0.2ex\hbox{/}}
\newcommand{\varepsdir}{\varepsilon\kern -5.5pt\raise 0.15ex\hbox{/}}

\newcommand{\pslash}{p \!\!\!/}

\newcommand{\quark}{\begin{picture}(300,100)(0,0)
\SetWidth{1.4}
\Line(50,50)(100,50)
\Line(100,50)(200,50)
\Line(200,50)(250,50)
\Line(67,40)(75,50)
\Line(67,60)(75,50)
\Line(142,40)(150,50)
\Line(142,60)(150,50)
\Line(217,40)(225,50)
\Line(217,60)(225,50)
\SetWidth{1}
\Vertex(100,50){2}
\Vertex(200,50){2}
\Gluon(100,50)(100,125){5}{3}
\Gluon(200,50)(200,125){5}{3}
\SetWidth{2}
\Line(105,130)(95,120)
\Line(105,120)(95,130)
\Line(205,130)(195,120)
\Line(205,120)(195,130)
\end{picture}}

\makeatother

\begin{document}
\thispagestyle{empty} 

\begin{flushright}
\begin{tabular}{l}
{\tt LPT Orsay 02-99}\\
{\tt UHU-FT/05-09}\\
\end{tabular}
\end{flushright}

\vskip 2.2cm\par
\begin{center}
\today\hskip 1 cm
version 2.0
\vskip 1cm
{\par\centering \textbf{\LARGE Artefacts and $<A^2>$ power corrections:} }\\
\vskip 0.2cm
{\par\centering \textbf{\LARGE revisiting the $MOM$ $Z_{\psi}(p^2)$ and $Z_V$} }\\

\vskip 0.9cm\par

{\par\centering \large  
\sc  Ph.~Boucaud$^a$,  F. de Soto$^b$
J.P.~Leroy$^a$, A.~Le~Yaouanc$^a$, J. Micheli$^a$, 
H.~Moutarde$^c$, O.~P\`ene$^a$, J.~Rodr\'{\i}guez--Quintero$^d$ }
{\par\centering \vskip 0.5 cm\par}
{\par\centering \textsl{ 
$^a$~Laboratoire de Physique Th\'eorique \footnote{Laboratoire associ\'e au CNRS, UMR 8627}(B\^at.210), Universit\'e de
Paris XI,\\ 
Centre d'Orsay, 91405 Orsay-Cedex, France.} \\
\vskip 0.3cm\par }{\par\centering \textsl{  
$^b$ Dpto. de F\'{\i}sica At\'omica, Molecular y Nuclear \\
Universidad de Sevilla, Apdo. 1065, 41080 Sevilla, Spain }\\
\vskip 0.3cm\par }{\par\centering \textsl{ 
$^c$ Centre de Physique Th\'eorique Ecole Polytechnique, 
91128 Palaiseau Cedex, France }\\
\vskip 0.3cm\par }{\par\centering \textsl{ 
 $^d$ Dpto. de F\'{\i}sica Aplicada e Ingenier\'{\i}a el\'ectrica \\
E.P.S. La R\'abida, Universidad de Huelva, 21819 Palos de la fra., Spain} 
\vskip 0.3cm\par }

  \end{center}

\vskip 0.45cm
\begin{abstract}
The NLO term in the OPE of the quark propagator vector part $Z_{\psi}$ 
and the  vertex function $g_1$ of the vector current 
in the Landau gauge should be dominated by  the
same $\langle A^2 \rangle$ condensate as in the gluon propagator. On the other hand, the perturbative part
has been calculated to a very high precision thanks to Chetyrkin and collaborators.
We test this on the lattice,
with both Wilson-clover and GW (Ginsparg-Wilson) overlap
fermion actions at $\beta=6.0,6.4,6.6,6.8$. 
Elucidation of discretisation artefacts appears to be absolutely crucial. 
First {\it hypercubic artefacts}
are eliminated by a powerful method. Other, 
very large, non perturbative, $O(4)$
symmetric artefacts, impede in general the analysis.
However, in two special cases with overlap
action - 1) for $Z_{\psi}$ ; 
2) for $g_1$, but only at large $p^2$ - 
we are able to identify the $\langle A^2 \rangle$ condensate; 
it agrees with the one
resulting from gluonic Green functions. We conclude
that the OPE analysis of quark and gluon Green function
has reached a quite consistent status, and that the power corrections have 
been correctly identified. A practical consequence of the whole analysis
is that the renormalisation constant $Z_{\psi}$ ($=Z_2^{-1}$ of the MOM scheme)
may differ sizeably from the one given by democratic selection methods.
More generally, the values of the renormalisation constants may be seriously
affected by the differences in the treatment of the various types
of artefacts, and by the subtraction of power corrections.

\end{abstract}
\vskip 0.4cm
{\small PACS: \sf 12.38.Gc  (Lattice QCD calculations)
\vskip 2.2 cm 
 
\setcounter{page}{1}
\setcounter{footnote}{0}
\setcounter{equation}{0}

\renewcommand{\thefootnote}{\arabic{footnote}}
\vspace*{-1.5cm}
\newpage
\setcounter{footnote}{0}
\section{Introduction}

The study of the quark propagator and vertex functions in momentum space has been  extensively
pursued in the literature starting in the 70's with analytical considerations  \cite{lane,pagels,politzer}, later completed by the discovery of the presence of a contribution of the $A^2$ operator, due to gauge fixing (\cite{lavelle1992}).
Numerical lattice QCD has more recently  treated this issue
\cite{bernard1990,pittori9411010,becirevic9909082,skullerud0102013,bowman0209129,dawson9909107,zhang0301018,bhattacharya0111002}. The scalar part of the quark propagator  is
related via axial Ward identities to the pseudoscalar vertex function. The role of the
Goldstone boson pole in the latter has been  thoroughly discussed
\cite{cudell9810058,cudell9909086,cudell0101009,becirevic0403044}.

In this paper we will mainly concentrate on the vector part of
the inverse quark propagator, the one which is proportional to $\pslash $, 
$Z_{\psi}$, and on the vector vertex function
$g_1$ related to it by the Ward identity, or, equivalently, on the ratio $Z_V^{MOM}(p)={Z_{\psi}(p^2) \over g_1(p^2)}$. 
One of the reason is to check for the effect of the $\langle A^2 \rangle$ condensate 
which has been discovered via power corrections 
to the gluon propagator and three point Green functions at large momenta ~\cite{boucaud0003020,boucaud0008043,boucaud0101302,desoto0105063}~(let us recall that, in order to perform trustable perturbative calculations, we take always $p>2.6~GeV$). OPE shows 
that the effect of the $\langle A^2 \rangle$ condensate should
have almost the same magnitude in the $Z_{\psi}(p^2)$ of the quark as in the gluon propagator. 
Moreover, the perturbative-QCD corrections 
are known to be varying especially slowly (the anomalous dimension being
zero at one loop in the Landau gauge), as we shall recall later. This should give
a favorable situation to display the power corrections
(recall that in the gluon case, it was difficult to
disentangle the power and the logarithmic corrections, which were moreover very sensitive to the value of $\Lambda_{QCD}$).
Now, for the Wilson-SW action, 
the crude values plotted in litterature (with only a selection of democratic points)
for $Z_\psi(p^2)$ are extremely flat above 2 GeV~\cite{becirevic9909039}
This was formerly considered as being natural, as a consequence of
the vanishing of one-loop anomalous dimension.  However, thinking more about it,   
it must be considered on the contrary as very worrying since it means the decrease predicted both by the perturbative QCD corrections
and the $\langle A^2 \rangle$ condensate {\it would not
be seen}. This lead us to start a very systematic study of the  
problem, with the following series of improvements  on  earlier
works:

\begin{itemize}
\item 
We reach an energy of 10~GeV by matching several lattice spacings 
so we are in  a better position to eliminate lattice artefacts and also to identify
power corrections, which requires a large momentum range.

\item 
We make use of a very efficient
way of eliminating {\it hypercubic} artefacts which we have first elaborated  while studying 
gluon propagators~\cite{boucaud9810322},~\cite{becirevic9910204}. 
In a recent paper \cite{boucaud0205187}, this method has still been 
improved for the specific case  of
the quark Green functions, where such artefacts are huge, especially in $Z_{\psi}$
and for the overlap action. Hypercubic artefacts have often been cured
by the ``democratic'' method which considers only momenta with equilibrated
values of the components. This method points  in the right direction but,
as we shall discuss  in more details, is by far insufficient  when the 
hypercubic artefacts reach such a level as illustrated in this article.

\item 
We make a systematic use of Ward-Takahashi identities relating the quark propagator
and the  vertex function. According to these
Ward identities, $Z_V^{MOM}$ should be independent of $p^2$ up to artefacts. Then, $Z_V^{MOM}$ is {\it a very sensitive test of the presence of artefacts}.
The fact that we observe 
a strong dependence of the
lattice $Z_V^{MOM}$ on $p^2$,  shows unambiguously the existence 
of large remaining discretisation
artefacts, this time respecting $O(4)$ symmetry, and decreasing with negative power of momentum. 
We can also check the consistency  of $Z_V^{MOM}$ with other determinations
of $Z_V$.

\item 
The above mentioned $O(4)$ symmetric artefacts could constitute a very serious problem, since we do not know how to eliminate them, and they would hide the OPE power corrections. Fortunately, in the case of overlap fermions (with $s=0$), we find that for $Z_{\psi}$, there remains only very small $O(4)$ artefacts. Then, we can obtain a satisfactory estimate of
the $1/p^2$ correction due to the $\langle A^2 \rangle$ condensate. 
For this reason, we consider mainly overlap fermions (with $s=0$),although they have some 
specific
unconveniencies (for example very large perturbative corrections).

\end{itemize}

This work, as well as the preceding one, clearly shows that 
lattice artefacts are overwhelming at the start, both hypercubic and $O(4)$ ones.
Hypercubic ones have been shown to be cleanly eliminated by our method. 
Remaining $O(4)$ ones, on the other hand,
cannot, and we do not foresee the possibility of a similarly efficient method,
wherefrom we have to rely on situations where they are small for reasons
which are not known, so that their smallness appears accidental.
In that respect, it may seem worrying ; but, on the other hand,
we are very happy to have found that the OPE can be checked to a 
good accuracy.

Another obvious interest of the study is then to improve the  determination of the MOM renormalisation constants, by taking into account both  continuum power corrections and artefacts ; and, indeed it indicates  that   much care must be exerted in using MOM renormalisation approach when high precision will be required, a point which has already been illustrated by the Goldstone contribution to $Z_P^{MOM}$.

In  section \ref{theor}, we will recall some theoretical premises, 
in section \ref{lattice} we will indicate the lattice conventions and the simulations
which we have performed, in section \ref{artihyp}  we will recall briefly the 
method to eliminate hypercubic lattice artefacts, 
in sections \ref{artiO(4)}, \ref{artivol}, we discuss other artefacts (Lorentz scalar artefacts, volume effects), in sections \ref{resultsoverlap}, \ref{ZAover}, \ref{resultsclover},\ref{gluon}
we will give the results and, in section \ref{concl},  
we will give our conclusions and further discussions.

\section{Theoretical premises}
\label{theor}

We work in the Landau gauge.
Let us first fix the notations that we will use.
We will use all along the Euclidean metric.  The continuum quark propagator
is a $12 \times 12$ matrix $S(p_\mu)$ for 3-color and 4-spinor indices. The inverse propagator is expanded according to :
\bea\label{Sm1}
S^{-1}(p) = \delta_{a,b} Z_\psi(p^2) \left( i\,\pslash + m(p^2)\right)
\eea
where $a,b$ are the color indices. $Z_\psi(p^2)$ being a standard lattice notation (for the precise lattice definition, see below, section \ref{lattice}).
Obviously, one has in the continuum, with trace on spin and color:
\bea\label{zpsi}
Z_\psi(p^2)=1/12~Tr(S(\mu)\gamma_{\mu} p_{\mu})/p^2 
\eea
Sometimes, one uses the alternative quantity:
\bea\label{b}
b(p^2) = Z_\psi(p^2) \, m(p^2)
\eea
to describe the scalar part of the propagator.

Let us consider a colorless vector current $\bar q \gamma_\mu q$.
 The three point Green function  $G_\mu$ is defined by
\bea\label{Gmu}
G_\mu(p, q) = \int d^4x d^4y~e^{i p \cdot y + i q \cdot x} 
< q(y)\bar q(x) \gamma_\mu q(x) \bar q(0) >
\eea

The vertex function is then defined by 
\bea\label{gammamu1}
\Gamma_\mu(p, q) = S^{-1}(p) \,G_\mu(p, q)\,S^{-1}(p+q)
\eea
In the whole  paper, we will restrict ourselves to the case where
the vector current carries a vanishing momentum transfer $q_\mu$. In the following we will omit to write $q_\mu=0$ and we will understand
$\Gamma_\mu(p_\mu)$ as the bare vertex function computed on the lattice.

From Lorentz covariance and discrete symmetries
\bea\label{gammamu2}
\Gamma_\mu(p) = \delta_{a,b} \left[g_1(p^2) \gamma_\mu + i g_2(p^2) p_\mu +
g_3(p^2) p_\mu \pslash + i g_4(p^2) [\gamma_\mu,\pslash]\right]
\eea
which should be obeyed approximately on the lattice, as we checked.

\subsection{Renormalisation and Ward-Takahashi Identities}

The renormalised vertex function is then $Z_2 Z_V \Gamma_\mu(p)$. Here, we must say something of conventions for renormalisation constants. The standard definition of renormalisation constants has been to {\it divide} the bare quantity by the renormalisation constant to obtain the renormalised quantity
(except for photon or gluon vertex renormalisations $Z_1$ which we do not use). $Z_2$ is the standard renormalisation of fermions $\psi_{bare}=\sqrt{Z_2} \psi_R$, $S(p)= Z_2(\mu^2) S_{\rm R}(p)$. In principle renormalisation of composite operators, for instance $Z_V$, should be defined similarly. We have followed this convention in our works on gluon fields. But, in the case of {\it quark} operators,
an opposite convention has become standard in lattice calculations : $(\bar \psi {\cal O} \psi)_{bare}=Z_{\cal O}^{-1} (\bar \psi {\cal O} \psi)_R$ ; we feel compelled to maintain this convention for the sake of comparison with parallel works on the lattice. This explains our writing of the renormalised vertex function. 
In the continuum $Z_V=1$ (conserved current). We keep $Z_V$ since the local 
vector current on the lattice is not conserved, and the discrepancy, which is of course an artefact, generates however finite effects in graphs due to additional divergencies multiplying the $a$ terms (which have higher dimension). 

The Ward identity in the renormalised form tells us that at infinite cutoff : 
\bea
(\Gamma_{\rm R})_\mu(p)=-i\,\frac {\partial}{\partial p^\mu}S_{\rm R}^{-1}(p)
\label{WTIR}
\eea
After multiplying both sides by  $Z_2^{-1}$ to return to bare quantities
\bea\label{ward1}
Z_V\Gamma_\mu(p) = -i\, \frac {\partial}{\partial p^\mu}S^{-1}(p),\quad
\eea
which from (\ref{Sm1})-(\ref{gammamu2}) implies
\bea\label{ward2}
Z_V g_1(p^2)=Z_\psi(p^2),\quad 
\,Z_V g_3(p^2)=2~\frac {\partial}{\partial p^2}Z_\psi(p^2),\quad 
- Z_V g_2(p^2)=2~\frac {\partial}{\partial p^2}b(p^2),\quad g_4(p^2)=0
\eea
We note that the first equation \ref{ward2} implies that $Z_V$ is independent of the renormalisation scheme {\it up to artefacts}(it is a ratio of bare quantities). Of course, this will hold up to terms vanishing as inverse powers of the cutoff at infinite cutoff, which are called {\it artefacts} in the lattice language. It must be recalled that, on the lattice, the Ward identity
is not exact, but holds only up to artefacts, because we work at finite cutoff, and the deviation will be found very large in some cases. A very important consequence
of the Ward identity for our study is that the ratio $g_1(p^2)/Z_\psi(p^2)$ is constant up to artefacts, or that deviations of this ratio from a constant are pure artefacts.

Defining analogously  the vertex function of the 
pseudoscalar ($\bar q \gamma_5 q$) density, 
\bea\label{gamma5}
\Gamma_5(p_\mu) = g_5(p^2) \gamma_5
\eea
the axial Ward identity implies
\bea \label{ward5}
  \frac{Z_P}{Z_S} m_q g_5(p^2) =  b(p^2) 
\eea 
where $m_q$ is $1/(2\,a)(1/\kappa-1/\kappa_c)$.


\subsection{MOM renormalisation ; radiative corrections} \label{radiatives}

To perform renormalisation on the lattice, we appeal as usual to the
convenient MOM schemes, which does not refer to a specific regularisation.
To speak technically, the precise renormalisation scheme 
that we use is the one called 
RI' by Chetyrkin, eq (26) in ref. \cite{Chetyrkin:hep-ph/9910332}. 
This is in fact the most standard MOM scheme
in the continuum, developed a long time ago by Georgi, Politzer, and Weinberg. 
It consists in setting the renormalised Green functions  to their {\it tree approximation} at the renormalisation point $p_i^2=\mu^2$, in the chiral limit. 
The inverse bare propagator 
$S^{-1}(p)$ is normalised through : 
\bea
S_{\rm R}^{-1}(p)|_{p^2=\mu^2} =
 \delta_{a,b} \left( i\,\pslash + m(p^2)\right)
\eea
Making $p^2=\mu^2$ shows that :
\bea
Z_2^{MOM}(p^2)=Z_\psi(p^2)^{-1}
\eea
up to artefacts. To renormalise the bare vertex function $g_1(p^2)$, 
we multiply it by the factor \\
$Z_2^{MOM}(\mu^2) Z_V^{MOM}(\mu^2)$. The MOM renormalisation
for $g_1$ must be chosen so that the renormalised W-T identity (eq. \ref{WTIR}) $g_1^{R}(p^2)=Z_\psi^{R}(p^2)$ holds, therefore $g_1^{\rm R}(p^2=\mu^2)=1$ ; we deduce that $Z_V^{MOM}(\mu^2)$ is the ratio $Z_{\psi}(\mu^2)/g_1(\mu^2)$ {\it up to artefacts}, but as remarked above,
this must be nothing but the scheme independent $Z_V$ : therefore this ratio $Z_V^{MOM}(\mu^2)$ is independent of $\mu^2$, {\it up to artefacts}. From now on, we define $Z_V^{MOM}$ as the ratio $Z_{\psi}(p^2)/g_1(p^2)$ (measured in fact on the lattice); we write :
\bea
Z_V^{MOM}(a^{-1},p^2)=Z_{\psi}(p^2)/g_1(p^2) \label{defzv}
\eea
which recalls that $Z_V^{MOM}$, which should be in fact {\it independent of $p^2$} in the limit where the cutoff $a^{-1}$ is infinite, is not so at finite $a^{-1}$-i.e. there are artefacts.

The Ward identity implies that $Z_\psi(p)$ and $g_1(p)$ have in particular the same
perturbative $p$ scale  dependence.
From the calculations of Chetyrkin et al.  
\cite{Chetyrkin:hep-ph/9910332}\cite{Chetyrkin:hep-ph/0008094}, we may express,  for example, the perturbative running of $Z_\psi$ at large $p$ as a function of the running $\alpha_{\rm MOM}(p)$. This is our main choice throughout this paper, although we discuss the effect of substituting an expansion in $\alpha_{\rm \overline {MS}}(p)$. More precisely, we will always choose to use the definition of $\alpha_{\rm MOM}(p)$ by the {\it symmetric} three-gluon vertex. The advantage of quarks is that we can reach an accuracy of four loops
in the RG expansion,
because we do not need $\beta_3^{MOM}$ (for symmetric $\alpha_{\rm MOM}(p)$), since the dimension of the fermion is 0
at lowest order in the Landau gauge.
Since the expression is lengthy and not necessary for present understanding, 
we refer the reader to the appendix \ref{A}. 
Of course, even with such accuracy, such an expression cannot be expected to hold for too small $p$ : we esteem the lower bound to be $p_{min}=2.6~GeV$, from our experience in the case of gluons ; indeed, we must avoid to go down too much close to the "bumps"
which manifest clearly that the gluon Green functions become non perturbative.
The perturbative calculation requires a value of $\Lambda_{QCD}$ ; one advantage
of  $Z_\psi(p)^2$ is that it is not so much sensitive to this value as the gluon quantities, because of the
vanishing of the $LO$ fermion anomalous dimension in the Landau gauge ; we choose $\Lambda_{\overline {MS}}=0.237~GeV$ from a previous analysis (\cite{boucaud0003020} ), not far
from the ALPHA estimate \cite{capitani-lat9810063},$\Lambda_{\overline {MS}}=0.238(19)~GeV$ ; we will discuss in the end the sensitivity of our results to this choice.

$Z_\psi(p^2)$ and $g_1(p^2)$ should have also the same non perturbative power corrections, up to a constant. We consider them now. 
\subsection{Power correction from the $<A^2>$ condensate} \label{power}
An OPE analysis as those performed in refs. \cite{boucaud0003020,boucaud0008043,boucaud0101302,desoto0105063} 
leads to consider a  $\langle A^2 \rangle$ condensate coupled to the quark propagator 
and vertex in Landau gauge. Let us recall that such a condensate could not
contribute to gauge invariant Green functions, and is present only in (gauge fixed)
gauge non invariant Green functions. The meaning and magnitude of such a condensate has been extensively discussed in the recent literature. Our aim 
here is to detect its effect on the Green functions through OPE, which provides
a way of testing theoretical ideas on its existence and magnitude.

For the propagator, we can write:

\bea\label{barecondensat}
        S^{-1}(p) = S^{-1}_{\rm pert}(p) + 
        i \pslash \ \frac{d_{\rm bare}\left(\alpha_{\rm bare}\right)}{p^2} 
        \frac{\langle : A^2_{\rm bare} : \rangle}{4 (N_C^2-1)} \ \delta_{a,b} + \cdots
\eea
where we only keep the leading term in $\pslash$. The calculation of the coefficients of the OPE has been performed in the chiral limit, and therefore one has as far as possible to stay near this limit.

In the renormalisation prescription  denoted by "~RI'~" in Chetyrkin papers (which amounts to the standard MOM of Georgi and Politzer in the chiral limit), and expanding everywhere in terms of $\alpha_s^{MOM}$, we obtain from Eq. (\ref{barecondensat})  
(see appendix \ref{B}):

\bea\label{condensat}
        \frac{Z_\psi(p^2)}{Z_\psi^{pert}(\mu^2)} = 
        \frac{Z_\psi^{pert}(p^2)}{Z_\psi^{pert}(\mu^2)} + \frac {32\pi}{3} \alpha(p)
        \left(\frac {\alpha(p)}{\alpha(\mu)}\right)^{-\frac 
        {\gamma_{A^2}^{(0)}-\gamma_0}{\beta_0}}
        \frac {<(A^2)_{\rm R} (\mu)>}{4 (N_C^2-1)} \frac 1 {p^2}
\eea
where $S^{-1}=i \pslash \ \delta_{a,b} \ Z_\psi$ and 
$S^{-1}_{\rm pert}=i \pslash \ \delta_{a,b} \ Z_\psi^{\rm pert}$ 
up to $O(m_q \pslash/p^2)$-terms. The condensate $<(A^2)_{\rm R} (\mu)>$ is renormalised
at the scale $\mu$. $Z_\psi^{pert}$ is given in eq.~(\ref{zpsi})
and the coefficients ($\gamma_0$ being the fermion anomalous dimension to lowest order) are
 \bea\label{gamma0}
\beta_0=11,\quad \gamma_0=0 \quad \gamma_{A^2}^{(0)} = \frac {35 N_c} {12}=\frac {35} {4}
\eea

As we have noted, $Z_V^{MOM}(p^2)$ should be constant in $p^2$  from the Ward identity (\ref{ward2}), up to artefacts ; then it cannot receive any power correction from $A^2$, and therefore,
$g_1$ receives exactly the same contribution from the condensate as $Z_\psi$.
We will  use  this as a very  useful test.

The essential step is then to fit this formula on the lattice data to extract 
$\langle A^2 \rangle$. The renormalisation constants at each $\beta$, ${Z_\psi(\mu^2,\beta)}$
will enter in the fit as free parameters to be determined, although they would be expected a priori to be close to lattice perturbation theory predictions. Of course, in general, we have to add lattice artefacts to eq.(\ref{condensat}), and one of the main problems we will discuss is how to determine them accurately.

An important warning must be made here, concerning the low accuracy in the perturbative calculation of the Wilson coefficient of $A^2$ written above : namely, it is only tree order with renormalisation group improvement. Expanding in terms of $\alpha_s^{MOM}$,
although it may seem natural, is completely arbitrary, and one would wish the results to be the same with $\alpha_s^{\overline {MS}}$. While this is the case to a good precision for $Z_\psi^{\rm pert}$, this is obviously not the case here,
due to the low order of the expansion : $\alpha_s(p)$ is quite different in the two schemes : at $p=2.6~GeV$, the ratio ${MOM}/{\overline {MS}}$ is around $2$ and decreases slowly down to $1.4$ at $10~GeV$ ; taking into account the anomalous dimension amounts roughly to replace $\alpha_s(p)$ by $\alpha_s(\mu)$ ; then  the ratio of coefficients in terms of the two coupling constants is only  slightly closer to 1 : it is 1.5 in average over the whole range . This means that the coefficient is reduced by $50 \%$ when using $\alpha_s^{\overline {MS}}$. This is due to the fact that ratio of coupling constants decreases only very slowly up to the largest available momenta. As a consequence, the determination of $\langle A^2 \rangle$
obtained by fitting the lattice data will be automatically affected by the same amount. We will give the results with the convention of using everywhere $\alpha_s^{MOM}$, as we have done for gluons. We shall first show that the power correction is indeed present and well determined, and then express it in terms of the condensate value, which suffers from the above uncertainty.
We also note that the ratio of condensates fitted from gluons
and quark Green functions, which should be 1 ideally, is not affected by this uncertainty, since the Wilson coefficients relative to the various Green functions differ mainly by purely algebraic numbers (the anomalous dimensions differ only slightly) \footnote{While finishing the article, we have become aware, thanks to D. Becirevic, of the calculation of the two-loop anomalous dimension of $A^2$ by \cite{dudal0308181}, in the ${\overline {MS}}$ scheme}.

\section{Lattice calculations}
\label{lattice}
We have first used SW-improved Wilson quarks (often called clover) with the
$c_{SW}$ coefficients computed in~\cite{luscher9609035}. 100 quenched gauge 
configurations have been computed at $\beta=6.0, 6.4, 6.6,6.8$ with
volumes $24^4$, $16^4$ and $8^4$. We have performed the  calculation
for five quark masses but in practice, for what is our concern in this
paper, the quark mass  dependence has not surprisingly proven to be
negligible ; anyway, since the theoretical calculations are performed in the chiral limit, we have to work as close as possible to the chiral limit ; then, we present only for simplicity the results for
the {\it lightest quark mass}, about $30~MeV$, {\it i.e.}
\bea
 \kappa= 0.1346,\,\,  0.13538,\,\,
0.13515,\,\,  0.13489 \quad {\rm for} \quad
\beta = 6.0,\,\, 6.4,\,\,6.6,\,\,6.8 
\eea 
 It should also be mentioned  
that all the results presented for clover action refer to the $24^4$ lattices unless stated otherwise. 

In addition to improved Wilson fermions, the use of overlap 
fermions~\cite{neuberger9707022} has revealed necessary,
and even crucial to obtain a good determination of the power correction. 
We have used approximately
the same  physical masses {\it i.e.} as in the improved Wilson case
\bea a\,m_0= 0.03,\,\,0.01667,\,\,0.01.\,\quad {\rm for} \quad
\beta = 6.0,\,\, 6.4,\,\,6.6,\,\,6.8  
\eea
with $s=0$ and volumes of $16^4$. 
 The bare mass $m_0$ and $s$ are defined from
\bea\label{overlap}
D_{\rm over}=  ( 1+s + am_0/2) + ( 1+s - am_0/2) 
\frac{D_{\rm w}(-(1+s))}
{\sqrt{D_{\rm w}(-(1+s))^\dagger \, D_{\rm w}(-(1+s))}}
\eea

where $ D_{\rm w}(-(1+s))$ is the Wilson-Dirac operator with a (negative)
mass term $-1-s$
\bea
  D_{\rm w}(-1-s) \equiv \frac{1}{2} \gamma_\mu ( \nabla_\mu +\nabla^*_\mu)
  - \frac{1}{2} a \nabla^*_\mu \nabla_\mu -1-s \,,
\eea

We use $s=0$. $s=0.4$ is considered preferable from locality requirements \cite{hernandezlat9808010}, however the difference is slight as soon as $\beta$
is larger than $6.0$.
The reason for using only the small lattice $16^4$ is well known ; it is due
to limitation in the special treatment needed for small eigenvalues of the Neuberger operator.
In practice, as for clover action, we discuss only the lighest quark mass,
roughly corresponding to the same $30~MeV$.
 
The propagators $S(x,0)$ from the origin to point $x$ have 
 been computed and their Fourier transform
\bea
\widetilde S(p) = \sum_{x} e^{-i p\cdot x} S(x,0)
\eea
have been averaged among all configurations and all momenta $p_\mu$ within
one orbit of the hypercubic symmetry group of the lattice, 
exactly as for gluon Green functions  
in~\cite{boucaud9810322}. 

In the case of overlap quarks 
the propagator and other Green functions are improved according to a standard 
and exact procedure~\cite{capitani9908029 } 
which should eliminate $O(a)$ discretization errors in Green functions, at large $p$, in the perturbative regime \footnote{Indeed, in our opinion, the argument on the vanishing of artefacts uses chiral symmetry of vacuum matrix elements, which holds only when spontaneous symmetry breaking can be neglected.}:
\bea\label{impr}
\widetilde S_\ast(p) = \frac {\widetilde S(p) - \frac 1 2}{1-am_0/2}
\eea
From now on, the notation $S(p)$ will represent the improved 
quark propagator in the case of overlap  quarks and the standard
 one in the case of clover quarks.

In both cases we fit the inverse quark propagator by
\bea
\widetilde S^{-1}(p) = \delta_{a,b}\, {Z_\psi(p)}\,\left(
 i\,\bar \pslash + m(p^2)\right)
\eea
according to eq.~(\ref{Sm1}) and where $\bar p_\mu$ is defined in 
eq. (\ref{tilbar}). We write $Z_\psi(p)$ because of the loss of the
Lorentz invariance.

$Z_\psi(p)$ can then be written as :
\bea
Z_\psi(p)=\frac {1} {12} {\rm Tr}\left[\gamma_\mu \bar p_\mu S^{-1}(p) \right]/(\bar p)^2
\eea
The three point Green functions
with vanishing momentum transfer are computed by averaging analogously over
the thermalised configurations and the points in each orbit
\bea
G_\mu(p, q=0)= <\gamma_5 \,\widetilde S(p)^\dagger \,\gamma_5 \,
\gamma_\mu \,\widetilde S(p)>.
\eea
where the identity $S(0,x)= \gamma_5 S^
\dagger(x,0) \gamma_5$ has been used.
The vertex function is then computed according to eq.~(\ref{gammamu1}) and 
we choose for the lattice form factor  $g_1$ :
\bea
g_1(p^2) = \frac {1} {36}{\rm Tr} \left[ \Gamma_\mu(p, q=0) 
\left(\gamma_\mu - \bar p_\mu \frac{\bar \pslash}{\bar p^2} 
\right)\right]
\eea
where the trace is understood over both color and Dirac indices.

Finally, according to the Ward identity (\ref{ward2}) we compute 
$Z_V^{MOM}$ simply from eq. (\ref{defzv})
where any effective $p^2$ dependence of $Z_V$ should come only from lattice artefacts.


Throughout this paper we will use the values in the following table
\ref{spacing} for the lattice
spacings, which follow the $\beta$ dependence found in ref.~\cite{capitani9810063}, appendix C, formule C.1,
\begin{table}[h]
\begin{center}
\begin{tabular}{||c||c|c|c|c||}
\hline
\hline
$\beta$ & 6.0 & 6.4 & 6.6 & 6.8 \\
\hline
$a^{-1}$ (GeV) & 1.966 & 3.66 & 4.744 & 6.1 \\
\hline
$a$ (fm) & 0.101 & 0.055 & 0.042 & 0.033 \\
\hline
\hline
\end{tabular}
\end{center}
\caption{{\small Lattices spacings}}\label{spacing}
\end{table}


\section{Elimination of hypercubic lattice artefacts}
\label{artihyp}
\subsection{Classification of artefacts}
The question of eliminating lattice artefacts has been 
perhaps our main difficulty in this work. 
Since we will have a detailed discussion, it is useful first to remind the main species of artefacts which are expected. 
First, we have {\it discretisation} artefacts \footnote{We will use the term "discretisation
artefact" preferably to the other common one, "ultraviolet artefact", because, as we shall find,
these artefacts may show up at small $p$ as well, due to non perturbative effects.}, which
themselves split into two : i) hypercubic artefacts, which are
the most visible because they break the elementary $SO(4)$ symmetry ;
they are seen, as we plot invariants of Green functions
as function of $p^2$, as  a large discrepancy between the value for different orbits at the same
$p^2$. With some simple treatment, it is easy to get one relatively regular
function of $p^2$. Nevertheless, in general, there remain non analytic
oscillations, and to eliminate them is both important to obtain the final
physical result, and demanding sophisticated methods.

ii) $SO(4)$ invariant discretisation artefacts, which remain after elimination of the cubic
ones,
and which will be discussed later. Let us say that this is the weakest point, because we do not
have theoretical principles to determine their form, neither there is a systematical empirical
method to determine them. 

Still, there may be finite volume artefacts which will be discussed also later, very
shortly since they do not seem sizable.

\subsection{Hypercubic artefacts}
\subsubsection{Generalities}

In successive papers \cite{becirevic9903364,boucaud0307026}, we have elaborated a very powerful method to deal with
hypercubic artefacts i.e. with those discretisation artefacts 
which come from the difference between the hypercubic 
geometry of the lattice and the fully hyper-spherically 
symmetric  one  of the continuum Euclidean space. The 
principle of this method \footnote{The initial idea is due to Claude Roiesnel} is based on identifying the 
artefacts which are invariant for the $H_4$ symmetry of 
the hypercube, but not for the $SO(4)$ symmetry of the
continuum.

Let us set the problem more precisely. Since we use hypercubic lattices our results are invariant under 
a discrete symmetry group, $H_4$, a  subgroup of the 
continuum Euclidean $SO(4)$, but not under $SO(4)$ itself. 
This implies  that lattice data
for momenta which are not related by an $H_4$ transformation
but are by a $SO(4)$ rotation will in principle differ.
Of course this difference must vanish 
when $a\to 0$ but it must be considered among the 
discretisation effects, i.e. ultraviolet artefacts. 
For example, in perturbative lattice calculations one encounters 
the expressions
\bea\label{tilbar}
\tilde p_\mu \equiv \frac 2 a \sin \left(\frac {a p_\mu} 2\right),\quad
\bar p_\mu \equiv \frac 1 a \sin \left( a p_\mu \right).
\eea
Both are equal to $p_\mu$ up to lattice artefacts: 
\bea\label{tilbar4}
\tilde {p^2} \equiv \sum_{\mu=1,4}\tilde p_\mu^2 = p^2 - \frac 1 {12} a^2 p^{[4]}
+ \cdots\,, \quad \bar p^2 =   p^2 - \frac 1 {3} a^2 p^{[4]}+ \cdots\,,\,{\rm
where}\, p^{[2n]} \equiv \sum_{\mu=1,4} p_\mu^{2n}
\eea

$p^2$, $\tilde p^2$, $\bar p^2$, $a^2 p^{[4]}$ are invariant under $H_4$
but only $p^2$ is under $SO(4)$. For example, the momenta $2\pi/L (1,1,1,1)$
and $ 2\pi/L (2,0,0,0)$ have the same $p^2$ but different $p^{[4]}, \tilde p^2$
and $\bar p^2$. In other words, if we call an orbit the set of momenta related by  $H_4$
transformations, different orbits correspond to the same $p^2$.  In general 
different orbits have different $p^{[4]}$.
The hypercubic artefacts can be detected,
considering a given quantity at a given $p^2$,  by looking carefully how it depends on the orbit.

These hypercubic artefacts, sometimes called ``anisotropy artefacts'',  
have been a long standing problem in lattice calculations, and of course, 
methods have been devised since a
long time to handle them. The general idea has been the so-called
"democratic" one  : the hypercubic effects are minimal when the four components
of the momentum for a given $p^2$ do not differ too much (this is democracy
between components ; ideal democracy is for diagonal $p\propto (1,1,1,1)$). 
Then the question is how to make
this criterion quantitative, the rationale being to find a compromise
between two contradictory requirements :
1) to be as much democratic as possible, which tends to reduce the number of points
2) to retain enough points to have a real curve.

The precise criterion is often something of a
secret recipe, not communicated in papers.
On the other hand , studying the gluon propagator, 
the authors of ref. \cite{leinweber9803015} have made explicit  a
selection method, keeping only the orbits having a point within a cylinder around the
diagonal. Several other similar criteria have been written.

\subsubsection{Our method :
the $\mathbf p^{[2n]}$ extrapolation method }

The alternative idea which we proposed, on the
contrary, relies on the use of all the orbits, and a method
to extract the physical point from an extrapolation of the different orbits.
A first successful application of this idea was for the gluon 
propagator~\cite{becirevic9903364}. We recall here the
final refined form of the method, the so-called $\mathbf p^{[2n]}$ extrapolation
method, presented in \cite{boucaud0307026}, and which has been shown to be necessary
to obtain satisfactory results for $Z_{\psi}$.

In order to perform a global fit we start from the remark that in
 this paper we are dealing with dimensionless quantities,
$g_1$ and $Z_\psi$. 
It is thus natural to expect that  hypercubic 
artefacts contribute via dimensionless quantities times a 
constant~\footnote{We neglect a possible logarithmic dependence on $p^2$.}.
Next we assume that there is a regular continuum
limit. We denote the generic Green function as Q, which depends a priori on
$p^2$, but also $a^2 p^{[4]}$, $a^4 p^{[6]}$ : 
$Q(p^2, a^2 p^{[4]}, a^4 p^{[6]}, a^4 p^{[4]},\cdots)$, 
and we Taylor expand it around
the $O(4)$ symmetric limit $Q(p^2,0,0,0,\cdots)$. 
Of course we must truncate this Taylor expansion of Q in $a$ and we choose to 
expand it up to $a^4$. Note that at this stage, the function $Q(p^2,0,0,0,\cdots)$
may still depend on $a$ through terms of the form $a^2 p^2$, etc..., but we do not consider
presently this further dependence.

The lattice results are $H_4$ invariant and thus typically functions of
$\tilde p^2$, $\bar p^2$ and $\tilde p\cdot \bar p \ $~\footnote{ In all 
this discussion we consider the mass as negligible.}. Then, let us consider a
typical $H4$ invariant and dimensionless term and expand it using eq. 
(\ref{tilbar4}):

\bea\label{check}
a^{2k} \frac{(\tilde p^2)^l (\bar p^2)^m (\tilde p\cdot \bar p)^n }  
{(\tilde p^2)^{l'} (\bar p^2)^{m'} (\tilde p\cdot \bar p)^{n'} } 
\rule[0cm]{8cm}{0cm}\nonumber \\
 = \ (a^2 p^2)^k \left(1 + v_1 \ a^2 \frac{p^{[4]}}{p^2} + \ ... \ +
v_i \prod_{j=1,N_i}a^{2 n_j} \frac{p^{[2 n_j+2]}}{p^2} + \ ...   \right) 
\eea
where $l+m+n-(l'+m'+n')=k$. In order to have a continuum limit $k \ge 0$.
As we expand the serie up to $a^4$ we have $k+\sum_{j} n_j \le 2$.
So for $k=0$ we obtain the following artefacts: $a^2 \frac{p^{[4]}}{p^2}\ $,
$\ a^4 (\frac{p^{[4]}}{p^2})^2\ $,$\ a^4 \frac{p^{[6]}}{p^2}$. For $k=1$ we 
have the only term $a^4 p^{[4]}$.
The coefficients $v_i$ could be
straightforwardly obtained in terms of $l,m,n,l',m',n'$.
 
As a conclusion, we have fitted  our results over the whole range of $p^2$ according to the following formula, where the $c_i$'s are constants independent of $p^2$ :

\bea\label{micheli2-1}
Q(p^2, a^2 p^{[4]},a^4 p^{[6]},\cdots) = Q(p^2, 0, 0) + 
c_1 \, \frac{a^2 p^{[4]}}{p^2}
+ c_2\,
\left(\frac{a^2 p^{[4]}}{p^2}\right)^2 + c_3\, \frac{a^4 p^{[6]}}{p^2} 
+ c_4\, a^4  p^{[4]}, \label{p2n}
\eea
with indeed small $\chi^2$'s. We have also checked the validity of this 
expansion for the free propagator.

The functional form used for $Q(p^2, 0, 0)$ does not influence significantly 
the resulting artifact coefficients. We can even avoid using any assumption 
about this functional form by taking all the values for $Q(p^2, 0, 0)$
as parameters which can be fitted~\footnote{We have enough data for that.}.

This improved correction of hypercubic artefacts turned out to be particularly 
necessary for $Z_\psi$. The results, including overlap-computed quantities, have already been presented in \cite{boucaud0307026}. The raw lattice data for $Z_\psi$ and $Z_V^{MOM}$ exhibit dramatically the ``half-fishbone'' structure which is a symptom
  of strong hypercubic artefacts, and we recall that these effects are especially strong in the overlap case. After applying the $\mathbf p^{[2n]}$ 
extrapolation method, eq. (\ref{p2n}), the curves
are now perfectly smooth ; they do not either exhibit the oscillations which remain in previous methods. 
We will return later to the fact that $Z_V^{MOM}$ is not at all a constant. 
   
Altogether we would like to recall the following hierarchy:
first, the hypercubic artefacts are one order of magnitude larger for 
overlap quarks  than for clover ones. Second, 
for both types of quarks the hypercubic artefacts
for $Z_\psi$ are one order of magnitude larger than those for 
$g_1$.

Let us stress that the discretisation artefacts we are discussing are all due to the {\it QCD interaction}. Indeed, our definition of $Z_{\psi}(p^2)$ is such that it is equal to 1, as in the continuum limit, when interaction is switched off, and we have also $g_1=1$ in that case. This illustrates that, in general,
it may not be sufficient, by far, to extract the free case artefacts, or to use such prescriptions as replacing $p_{\mu}$ by $\bar p_{\mu}$.

\subsubsection{Quantitative comparison with the ``democratic''  method}

We would like also to recall the quantitative comparison of our  `` $\mathbf p^{[2n]}$
extrapolation method'', eq.  (\ref{micheli2-1}) with the more common "democratic
selection" methods ;  the latter method is carefully defined in \cite{leinweber9803015}. This comparison is important, since almost all works up to now are using some variant of the democratic method, and since the difference with this method is crucial, as we show, to extract power corrections.

Let us consider the overlap case. If we try to select, \cite{leinweber9803015}, the orbits which are in
a cylinder around the diagonal with a radius $2\pi/L$, this is too restrictive anyway for our  $16^4$ overlap case, where we have only 22 orbits at the start. In  order to  have a less restrictive democratic criterion and to make a  bridge with our own method, we  will use the $p^{[2n]}$'s defined in eq. (\ref{tilbar4}).
In our language, democracy can be translated  as having a small enough ratio
$p^{[4]}/(p^2)^2$. 
Momenta proportional to $(1,1,1,1)$ and $(1,0,0,0)$ have ratios $1/4$ (minimum ratio, maximally democratic) 
and $1$ (maximum, totally undemocratic) respectively. We then retain the ``democratic'' orbits defined  by an intermediate
$p^{[4]}/(p^2)^2 \le .5$. This leaves already only 7 orbits out of 22 for every $\beta$.
 
In fig. \ref{qZdemos} we plot for $Z_\psi$ the result 
of this selection, as compared with our own method. 
Fig. \ref{qZdemos} clearly  shows visible oscillations in the democratic curve demonstrating
that the hypercubic artefacts have not been totally eliminated, while our treatment yields something perfectly regular. 
We prefer our own method for this reason and also because of the
loss of information due to the rejection of ``undemocratic'' points, which leaves us with very few points. This appears
crucial in the overlap case, where one is constrained to use small lattices.

Then another very important aspect appears : while completely
eliminating the hypercubic oscillations, we also considerably modify the {\it mean 
value of the curve}, as defined by an analytic fit ; for the present case, as seen on the figure, our curve is {\it considerably lower, with a quite steeper descent} than the democratic one.  
\vskip 0.5 cm
\begin{figure}[hbt]
\begin{center}
\leavevmode
\mbox{\epsfig{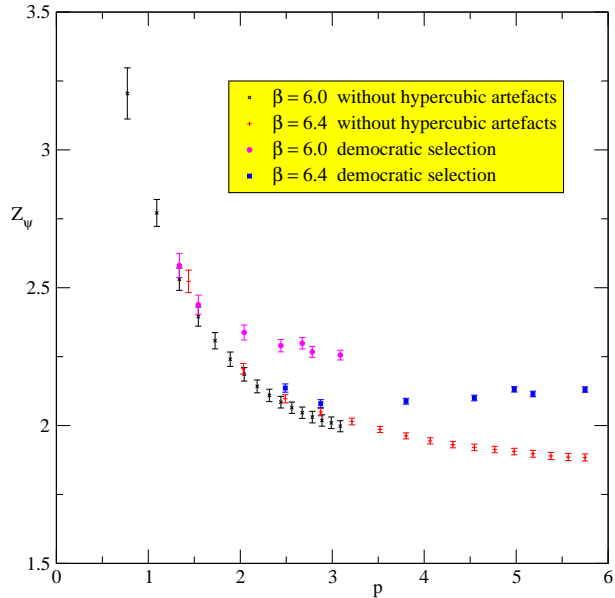}}
\vskip -0.5 cm
\caption{\small Comparison between the "democratic" method and our refined treatment   {} for the overlap $Z_\psi(p^2)$ at $6.0$ and $6.4$. Above 2 GeV, precisely where one must  study the power corrections, the difference is striking.}
\label{qZdemos}
\end{center}
\end{figure}

\subsubsection{The importance of optimising the elimination of
hypercubic artefacts}

Let us stress then that, as is particularly visible from this comparison, the difference of the $\mathbf a^2 p^{[2n]}$ method 
with the standard democratic
method is not at all academic in the context of the study of power corrections and renormalisation constants.
The difference with the previous versions of our method is also not negligible,
as we have found.

{\bf 1)} It is obvious that with the standard democratic method, we would obtain quite different
results for power corrections and $O(4)$-symmetric artefacts, and therefore for the 
resulting perturbative contribution. In fact, it has not even been really considered
that $Z_{\psi}$ could be affected by such power corrections.
{\bf 2)}Moreover, we observe
that the power corrections, as well as the residual  
$O(4)$ symmetric discretization artefacts,
extracted by the previous variants  of our method for treating 
the hypercubic artefacts 
are not the same ; indeed, when using a 
previous cruder treatment
for overlap action,
we have found an important $a^2 p^2$ artefact, which disappears
with the more refined treatment,
and we were also finding different power corrections (with a weaker condensate).
The Wilson case shows similar spuriosities.
This means that for a too crude treatment, 
some hypercubic artefacts can be spuriously mimicked as 
part of $O(4)$ symmetric discretization artefacts or continuum power
corrections.
This {\it does not} imply that the determination 
of power corrections is uncertain
in this respect, but rather that it is very important 
to push hypercubic artefact elimination to the
best to obtain the genuine continuum power corrections.

Let us finally mention an interesting consequence ;
as is well known, the values of Green functions at different momentum points in a Monte-Carlo lattice calculation are highly correlated, which should
lead to very small $\chi^2/{d.o.f.}$ for fits describing the
$p$ dependence by smooth analytic expressions (by small, we mean well below one). It is not found so
with too crude treatments of the hypercubic artefacts, 
because of the erratic oscillations which always remain in the latter methods
mimicking statistical deviations.
We observe however an impressive decrease of the $\chi^2$ 
down to its expected small value when we 
improve the treatment of the data, showing that we are now obtaining
indeed very smooth functions as physically expected. 

\section{Proof of the presence of "non canonical" artefacts : $Z_V^{MOM}$ and $O(4)$ symmetric discretization artefacts. Problems for the
determination of the continuum limit.} \label{artiO(4)}

We now start, in the rest of the discussion, from the data obtained
through the above treatment of the hypercubic artefacts. They still differ from
the continuum by renormalisation and by $O(4)$ symmetric discretization artefacts.
It happens that the determination of these artefacts is still harder in general
than for hypercubic ones.
(as for finite volume artefacts, which we estimate to be weak,
see the corresponding short section below).

Indeed, in practice, there is no similar
unambiguous method to determine the $O(4)$ symmetric discretization artefacts.
True, they manifest themselves by a residual variation with $\beta$, and, in
principle, we could study the variation with  $\beta$ at each momentum, and then
extrapolate to the continuum. However, 
this requires too many momenta and too much
accuracy, if we want really to extract the power correction from the
extrapolation to the continuum. 

{\bf a)} a general method for treating $O(4)$ scalar artefacts 

Then, a more practical method 
consists in assuming
a prescribed analytical form for both the continuum and the artefacts, with some
unknown parameters to be determined by $\chi^2$ adjustment.

One has therefore to appeal to our a priori 
knowledge
a) of the continuum, as  function  of $p$ ; b) of the structure of $O(4)$ 
symmetric artefacts, 
as  function  of $p$ and $a$, so that we could make fits with prescribed 
functions
depending on a limited number of free parameters. 
For the continuum, this is exactly what is
provided by the OPE, with the renormalisation constants $z_b$ and $\langle A^2 \rangle$ 
as free parameters.
For $O(4)$ symmetric artefacts, a standard  idea is to recourse to 
{\it lattice perturbation theory}, in which the structure 
of artefacts is easily
explicited. After all, this is what we have invoked 
for the hypercubic artefacts.
As for $O(4)$ symmetric artefacts, the result is quite simple : 
in the case of a scalar function, and in the chiral limit, 
where there
is no other dimensioned parameters than $a$ and $p$, 
there can be no other artefacts
than $a^n p^n$, with $n>0$. Then we could work with a few parameters only.

{\bf b)} why it fails

But now, there appears a very unfortunate circumstance. 
It is seen that this
usual assumption  of a perturbative structure of artefacts 
does not work at all, at least in
general. We observe undoubtedly $O(4)$ symmetric artefacts {\it decreasing with p},
i.e. for instance of the form $1/p^n$ and $n>0$ times some positive power of $a$.
To show it, the best way  is to consider
the $p$ dependence of $Z_V^{MOM}=g_1/Z_{\psi}$, which
should be momentum independent close to the continuum limit. Any $p$ dependence
is therefore  to be attributed to artefacts. 
Now, the lattice data show quite clearly
a very strong $p$ dependence of $Z_V^{MOM}$ except at large $p$. This is true for clover
action, and for overlap action as well, see  Fig.  \ref {qZVoverlap} for overlap fermions.\par

\vskip 1 cm

\begin{figure}[h!]
\begin{center}
\leavevmode
\mbox{\epsfig{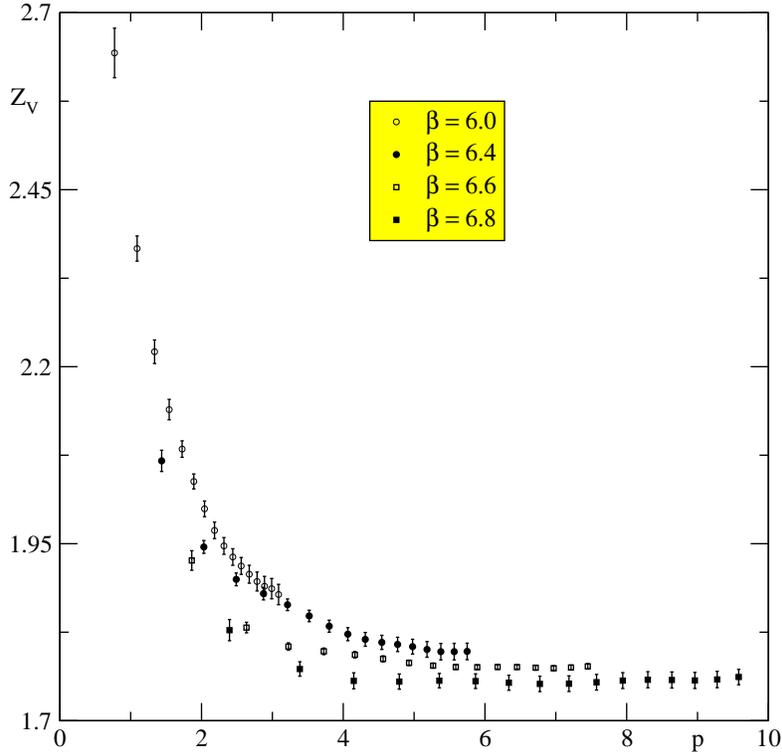}}

\vskip -0.5 cm
\caption{\small Overlap $Z_V^{MOM}$ as function of momentum, respectively for $\beta=6.0$ and $6.4$, and $\beta=6.6$ and $6.8$, for the overlap action. The strong momentum dependence which is displayed must be entirely attributed to artefacts.}
\label{qZVoverlap}
\end{center}
\end{figure}

The 
artefacts are similar ; they decrease monotoneously with increasing $p$ for
$\beta=6.4,6.6,6.8$; they
could be fitted by negative powers of the momentum squared ; for the overlap case,
this is also true for $\beta=6.0$ .

{\bf c)} consequences for determination of power corrections

The necessity of taking into account such negative powers of $p$
is very embarrassing, because, as we explained, we cannot distinguish
the $O(4)$ symmetric discretization artefacts by the sole  $\beta$
dependence ; we have to rely on the $p$ dependence. Now, it is clear 
that we will have much difficulty in distinguishing
the power corrections and the artefacts decreasing with $p$,
and therefore to determinate the condensate. One of the problems is that when increasing the number of negative power terms in the description of the artefact, the tendency is to get alternating signs, and therefore rather unstable results. There is no rationale as to where we should stop. In addition, we must
consider that we have also as parameters in the fit the renormalisation
constants $z_b$, which are practically  chosen as an independent parameter at each $\beta$. 
To add to the uncertainty, we observe that within the
precision of the data, we can obtain equivalently good solutions
by modifying the $O(4)$ artefacts with a correlative
change in the $z_b$'s.
Then, in spite of many efforts, we have in fact not been able to extract
{\it stable} and accurate values of $\langle A^2 \rangle$ by this method, although we have got a clear signal
that it is {\it positive and sizable}. Then, we have been able to fix
the continuum power corrections only by exploiting particular
circumstances, to which we devote the rest of the analysis, after a few words
on finite volume artefacts.

\section{Finite volume artefacts} \label{artivol}

Let us recall that in the Wilson case \cite{boucaud0307026}, we have found only very small
volume artefacts, after a careful study with $8^4, 16^4, 24 ^4$ lattices.
Only the first points with smallest $n^2$ were showing some effect.

We have not performed the same tests for the
overlap action, because of the slowness of numerical calculation ; 
we have only ran on a $16^4$ lattice. We are
conscious that this is a possible weak point, since this volume is small, and 
we rely mainly on the analysis of overlap results for the determination
of the condensate. So we think of extending this analysis to
larger volume as soon as possible. 

We stress however that the consistency which is obtained for the continuum power corrections between the overlap
results and the S-W (clover) ones with a greater volume $24^4$ (see section \ref{resultsclover}), is a further proof that the volume effects are not crucial for our purpose of determining the {\it condensate}.

We must also observe that we can determine the condensate mainly from our
large $p$ data ($p>4-5~GeV$)(see \ref{petitefenetre}), with large $L p$, where volume artefacts
are expected to be even smaller.

\subsection{Discretisation or volume artefacts ?}

The fact that the above $O(4)$ symmetric artefacts are observed at small $p$ -and only at small $p$
is deserving a special discussion, since it is so counter to usual expectation. Our ears are indeed accustomed to the dictum :
"artefacts at small $p$, finite volume artefacts", 
"artefacts at large $p$, finite spacing artefacts". Why are they not volume artefacts ? Since it is one
important finding of our study, we collect here our arguments.

Let us stress that, in the overlap case, the use of a
small volume $16^4$ at large $\beta$, to which we are constrained,
is not the reason for our finding of large $O(4)$ symmetric artefacts at small $p$. The first argument is that they are also seen in the clover case where we have tested
in detail the smallness of volume artefacts; to reiterate, we have seen that only one or two of the smallest momenta seem affected by a volume dependence,
while the artefacts we are discussing now are present over a large number of points .

Moreover, let us stress that
they have the typical behavior
of discretisation artefacts, i.e. they {\it decrease}  at large $\beta$ and fixed $p$ : indeed, $Z_V$
(which should be flat in the continuum limit) is flatter and flatter as $beta$
increases at fixed number of sites $16^4$, i.e. as the physical volume decreases. More precisely, we observe that, as the volume decreases, we have more and more points
where it is flat, i.e. where it has small artefacts : the respective number of points is  $0$ at $\beta=6.0$, $3$ at $\beta=6.4$,$9$ at $\beta=6.6$, $14$ at $\beta=6.8$. This is exactly counter to a volume effect, for which the flatness should be obtained for $n^2$ larger that some {\it fixed number}. In other terms, for a volume effect,
we would expect the artefacts to vanish at smaller and smaller values of $p$
as $\beta$ decreases (i.e. when the physical volume increases). Instead, we find that they vanish around $4-5~GeV$ irrespective of $\beta=6.4,6.6,6.8$ ; even more, at $\beta=6.0$, with the largest volume, $Z_V$ never becomes flat .

On the other hand, we have no understanding of why $Z_V^{MOM}$ seems devoid of sizeable
artefacts at large $p$ : it must be regarded as an accident.  But it must be underlined that this is not an accident specific to our particular problem.
It is a well-known fact, on which
any MOM practitioner is relying without being able to explain it : one measures
the $Z$'s in regions of large $p$, assuming that artefacts are small,
although they would be expected to be large precisely there from lattice perturbation theory.

\section{Cases with small $O(4)$ symmetric artefacts
with the overlap action at $s=0$. Determination of
the condensate.} \label{resultsoverlap}
As we have underlined as conclusion of section \ref{artiO(4)}, paragraph c),
the $O(4)$ symmetric artefacts we have found, when present, impede a clear
determination of the power correction. It is fortunate that they seem absent
or small in some cases. All correspond to overlap action ; then the results are convincingly consistent. Therefore, we concentrate on the overlap fermions in the present and following sections
(\ref{resultsoverlap},\ref{ZAover}). As concerns the clover action, we will not be able  to have so compelling conclusions, but we show that they are at least compatible (see section \ref{resultsclover}).

\subsection{Large $p$ ($p\sim or >5~GeV$)} \label{petitefenetre}
At this point, we notice that the above difficulties
may disappear first at large $p$, above roughly $5~GeV$ 
(therefore at $\beta=6.4,6.6,6.8$), 
because $Z_V$ is constant in this region, especially 
for the overlap action ; therefore it is suggestive
that the artefacts of $Z_{\psi}$ and $g_1$ are small there,
barring for the unprobable eventuality that $Z_{\psi}$ and $g_1$ 
would happen to have exactly the
same artefacts. We then fit $Z_{\psi}$ and $g_1$ with a formula
containing no $O(4)$ artefact,. i.e. we take only the continuum expression
times the renormalisation factor $z_b$, at fixed $\beta$, and we obtain then
a very encouraging conclusion. Choosing the window of $p$ within which
$Z_V$ is well constant, we obtain at each $\beta=6.4,6.6,6.8$, 
and for $Z_{\psi}$ and $g_1$, i.e. for {\it six} independent data,
almost the same condensate value ; we quote a common fit (see the Fig \ref{qfitA2petitefenetre}) to the three $Z_{\psi}$, with $p_{min}=5,5,4~GeV$ respectively, corresponding to the respective $p$ where $Z_V$ is beginning to be flat  :

\bea
              \langle A^2 \rangle \simeq (3.1 \pm 0.3)~GeV^2 \label{condensateI}
\eea	      
large and positive, with a rather small error, and quite consistent with what is found in the gluon
sector (see below). The $\chi^2/d.o.f.=0.07$ is very small as expected from very correlated data. Let us recall that this value of $\langle A^2 \rangle$ is obtained with the convention that the Wilson coefficient of the operator is expressed in terms of $\alpha_s^{MOM}$.
\vskip 0.5 cm
\begin{figure}[hbt]
\begin{center}
\leavevmode
\mbox{\epsfig{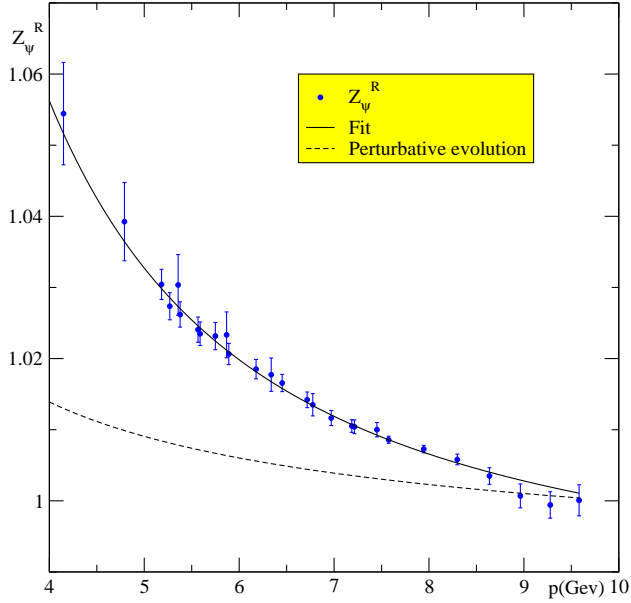}}

\vskip -0.5 cm
\caption{\small Renormalised overlap $Z_{\psi}$ at $\beta=6.4,6.6,6.8$ and large $p$, compared with the purely perturbative result .
The fit of the solid line is made with the perturbative part and only the condensate term , in the windows where the respective $Z_V^{MOM}$
become flat (solid line). One finds $\langle A^2 \rangle=3.1 \pm 0.3$. Note that $Z_{\psi}$ is still far from being flat. It has a monotoneous decrease, of around $6\%$ over the range, mainly due to the condensate. The dotted line is the purely perturbative result, clearly inconsistent with the data. Normalisation is made at $10~GeV$. $g_1$ is described by a similar fit. }
\label{qfitA2petitefenetre}
\end{center}
\end{figure}

A remarkable feature of this region of momentum is that both
$Z_{\psi}$ and $g_1$, and therefore $Z_V$ too, are almost independent
of $\beta$. We take it as a further indication
that artefacts are accidentally small there.

Let us reinsist that the value of the condensate given in eq. (\ref{condensateI})
corresponds to the convention, always followed in this paper, that the Wilson coefficient of the operator, calculated only at leading order, is expressed in terms of $\alpha^{MOM}$. The choice of $\alpha^{\overline {MS}}$, would lead to an appreciably higher value (larger by around $70\%$). However, what is important is that the power correction by itself is well
determined by our analysis, almost independently of such a change. Indeed, let us replace the OPE expression for the power correction by a simple power, without logarithms corresponding to $\alpha_s$, while maintaining the full perturbative expressions for the perturbative part. The fit then gives for the coefficient $c$ of the power term $c/p^2$:
\bea
c= (0.767 \pm    0.083)~GeV^2  \label{cMOM}
\eea
when using $\alpha^{MOM}$ in the perturbative part, and :
\bea
c= (0.844  \pm    0.083)~GeV^2 
\eea
a small change indeed, of only $10\%$, reflecting the small change of the perturbative
part, which is calculated by theory to a great accuracy.

\subsection{$Z_{\psi}$ over the whole allowed range of $p$}\label{grande fenetre}
In addition,
we observe another fact : $Z_{\psi}$ - but not $g_1$ - is  strikingly independent
of $\beta$ over the full range of $p$. We have no explanation for that, but we can at least interpret this as meaning that $Z_{\psi}$ is
free of $O(4)$-symmetric artefacts over all this range. 
And indeed, we can fit $Z_{\psi}$ on the full range  
$p>2.6~GeV$ allowed for the perturbative calculation, 
and the four $\beta$'s with :
\bea
                    \langle A^2 \rangle=(2.73 \pm  0.21)~GeV^2 
\eea		    
See the figure \ref{qfitA2grandefenetre}.
\vskip 0.5 cm
\begin{figure}[hbt]
\begin{center}
\leavevmode
\mbox{\epsfig{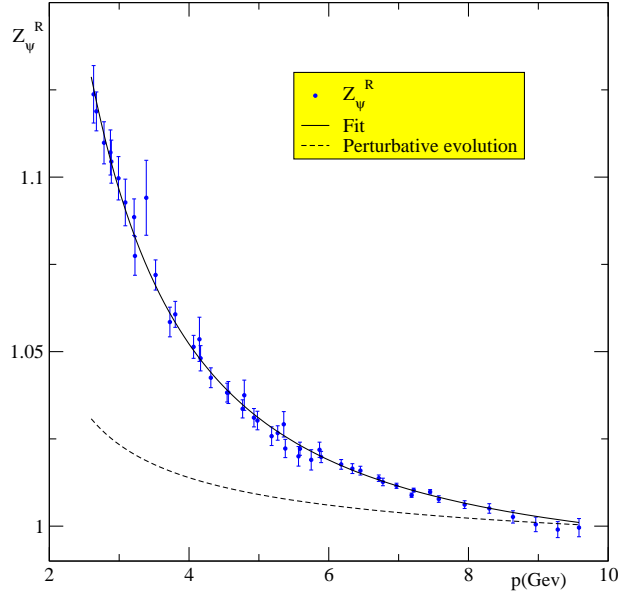}}
\vskip -0.5 cm
\caption{\small Overlap $Z_{\psi}$ at $\beta=6.0,6.4,6.6,6.8$, renormalised at $10~GeV$. The fit for $Z_{\psi}$ extends to the full allowed range of momentum, with $\beta=6.0$ included, and a close value of the condensate $\langle A^2 \rangle=2.75 \pm 0.2$. The dotted line is the purely perturbative result.}
\label{qfitA2grandefenetre}
\end{center}
\end{figure}

We thus obtain a remarkable similarity of the condensate with the previous value. We also obtain similar results by varying the window
of the momenta, provided the lower limit is not pushed beyond $5~$ GeV, and also
by selecting various triplets of $\beta$ values. This seems to support the consistency of our assumptions about artefacts.

If we want to still improve the agreement with the large $p$ analysis, we can introduce a further $1/p^4$ term, accounting for the possibility that we may not be sufficiently asymptotic to have a good description with the $1/p^2$ condensate alone. For the full window, and the
four $\beta$'s, we obtain a very good fit with :
\bea
                    \langle A^2 \rangle=(3.2 \pm 0.3)~GeV^2 
\eea

and with a small subleading term:
\bea
 		   (- 1.1 \pm 0.4)~GeV^4/p^4
\eea
with a minus sign which explains that the value of $\langle A^2 \rangle$ is slightly higher than
in the preceding fits. It is then closer to the large $p$ fit.

\subsection{$Z_V^{MOM}$}
From our analysis, we can deduce values of $Z_V^{MOM}$ standardly defined, i.e. free from artefacts :
\bea
  &&Z_V^{MOM,~large~p}(6.4)= 1.798 \pm 0.012, \nonumber\\ 
 &&Z_V^{MOM,~large ~p}(6.6)= 1.776 \pm 0.003,~~Z_V^{MOM,~large ~p}(6.8)=1.756 \pm 0.011 \label{ZVnum}
\eea
"$large~p$" superscript being to recall that we have selected the value at large p, where we hope to have small $O(4)$ artefacts, in view of the observed flatness ; the precise value is obtained by a fit.
For $\beta=6.0$, we quote $Z_V^{MOM,~large~p}(6.0)=  1.878 \pm 0.014$, but $Z_V^{MOM}$ is not yet flat at the highest momentum. Let us repeat that it is remarkable, and perhaps surprising, to observe such a constancy with $\beta$,
while one loop perturbation theory would predict a strong variation with $\beta$.  

\subsection{Comparison with lattice perturbation theory}
The fact that $Z_{\psi}$ and $Z_V$ are very different from $1$ (in fact not far from $2.$) in the overlap case with $s=0$ may seem surprising. But in fact, already in lattice perturbation theory, the tendency is that the one-loop corrections to $Z_{\psi}$ and $Z_V$ are large, because of a very large tadpole contribution to the self energy \cite{lat0002010,giusti-lat0007011} (while $g_1$ remains close to $1$ as we find non pertubatively). The net effect is already large at $\beta=6.0$,for the usual $s=0.4$ : $Z_V=1.247$, and still larger for our $s=0$ : $Z_V=1.444$ ; we use Table 1 of ref. \cite{giusti-lat0007011}  for the analytical expressions (the definition of $Z_{\psi}$ is different, but by a negligible amount); the numbers are quoted assuming a $BPT$ boosted coupling with $g_{BPT}^2=1.68$ (see ref. \cite{lat0002010} under eq. (44)). Of course, what is surprising is  first that the non perturbative determination (\ref{ZVnum}) is still  much larger,
and, second, that it is almost independent of $\beta$ over a large range. 

Note that the value found by ref. \cite{giusti-lat0110184} for $Z_A$ at $\beta=6.0$,$Z_A=1.55$, which should equate $Z_V$, is also much larger than BPT ; however, this is not $Z_A^{MOM}$ as measured directly ; it is the $Z_A$ deduced from hadronic W-T identities. One may think of large artefacts which render different the results from various definitions of $Z_A,Z_V$ ; our study below, section \ref{ZA-ZV}, shows indeed the presence of such effects,
very large at $\beta=6.0$; but they are decreasing rapidly at larger $\beta$, and our finding is that, at $\beta=6.8$, $Z_V^{MOM}$ is still much larger than the boosted lattice perturbation theory.

\section{$Z_V$ and $Z_A$. Consistency checks of the overlap action results. Complementary studies on 
chiral symmetry and artefacts } \label{ZAover}
To ascertain the soundness of our analysis, which may be surprising in several respects, we also perform several consistency checks of general nature in the overlap case ; indeed,
some strong statements deriving from chiral symmetry can be formulated.  

\subsection{$Z_A/Z_V$ in the non perturbative MOM scheme}
One should expect $Z_A^{MOM}/Z_V^{MOM}$ for the improved Green functions
to be exactly 1, i.e. without any artefact in the chiral limit, and in the perturbative regime, i.e at large momentum. 
This derives from the exact chiral symmetry of the action, discovered by Luescher,combined with the choice of the improved current, or equivalently,
of improved Green functions. However, one has also to assume the symmetry of the vacuum state, therefore the absence of spontaneous symmetry breaking. Therefore this result only holds at large momenta. We find indeed this to a high accuracy, fig. \ref{qZAsurZV}. Above $2~GeV$, for the four $\beta$'s, we have $Z_A^{MOM}/Z_V^{MOM}=1$ to a very high accuracy. On the other
hand, we see that for lower momentum i) the ratio differs from 1, and ii) it depends on $\beta$. The change of regime is rather abrupt. Observation i) has also been made for domain wall fermions at $\beta=6.0$ \cite{dawson0011036}.
\vskip 0.5 cm
\begin{figure}[hbt]
\begin{center}
\leavevmode
\mbox{\epsfig{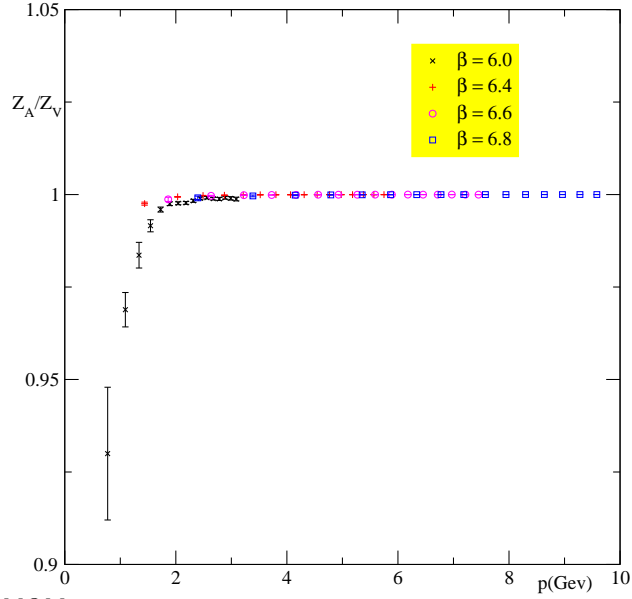}}
\vskip -0.5 cm
\caption{\small $Z_A^{MOM}/Z_V^{MOM}$ for overlap fermions and small quark mass at the four $\beta=6.0,6.4,6.6,6.8$. The ratio is 1 above $2~GeV$, but differs from 1 and depends on $\beta$ below $2~GeV$.}
\label{qZAsurZV}
\end{center}
\end{figure}
 	      
\subsection{$Z_A$ from a hadronic W-T identity against $Z_V^{MOM}$} 
\label{ZA-ZV}

In principle, once artefacts and power corrections have been eliminated, the 
various ways of defining the renormalisation constants can be related by perturbation theory.  In the case of $Z_V$ or similar cases, for the same action, implying identical finite parts, they should even be equal. Moreover, $Z_A$ should be equal to $Z_V$ from the chiral symmetry of the overlap action.\par 

We verify this statement, as a highly non trivial consistency check of our treatment, by measuring $Z_A$ from a standard hadronic Ward-Takayashi identity. $Z_A^{WI}=m_q/\rho$, where :
\bea
\rho=1/2 <\partial_0 A_0,P_5 >/<P_5,P_5>
\eea
In contrast to $Z_V^{MOM,large~p}$, $Z_A^{WI}$ presents a very strong variation with $\beta$ : at $6.0$, the situation seems hopeless, with $Z_A^{WI}\sim 3$, but the decrease is rapid :
\bea
Z_A^{WI}(6.0)= 3.03, Z_A^{WI}(6.4)= 2.046, Z_A^{WI}(6.6)=1.90, Z_A^{WI}(6.8)=1.78
\eea
and one reaches finally a level close to $Z_V^{MOM,~large~p}$. Moreover,
$Z_A^{WI}(a)$ is linear in $a^2$ to a very good precision :
\bea
Z_A^{WI}(a)=1.65+5.52 a^2,
\eea
which shows that the difference $Z_A^{WI}-Z_V^{MOM,large~p}$ is indeed a discretisation artefact, 
as it should, and one of the most canonical species, since we expect precisely
chiral symmetry breaking to be at most ${\cal O}(a^2)$. Indeed, on hadron states, the Ward identity is valid up to ${\cal O}(a^2)$, and Green functions at large $p$, where we measure $Z_V^{MOM,large~p}$, have also only ${\cal O}(a^2)$ artefacts from chiral symmetry arguments, in the chiral limit.

\section{Check through the Wilson clover $Z_{\psi}$} \label{resultsclover}
In view of the rather paradoxical situation which allows us to determine $\langle A^2 \rangle$
from the $s=0$ overlap action, it is important to reascertain this analysis
by a study of the standard clover action, to check whether we can obtain a similar {\it
continuum} result. Let us stress indeed that from renormalisability, the whole continuum function
must be the same up to an overall constant for all the versions of the action.
A priori, as we have said, the Wilson case may seem hopeless because there are large $O(4)$ artefacts even in $Z_{\psi}$ ;  and even at large $p$, $Z_V$ is not so flat

However, one can benefit from the knowledge gained in the overlap case. In the end, 
the situation appears in fact 
somewhat similar for the clover action as concerns $O(4)$ symmetric artefacts: 
the main burden of non canonical artefacts appears to concern $g_1$,
and we can obtain a good description of $Z_{\psi}$ with a minimal $a^2 p^2$
{\it canonical} artefact, and just one $a/p^2$ term. Indeed, we obtain :
\bea
\langle A^2 \rangle=(2.4 \pm 0.3)~GeV^2 
\eea

with artefacts
\bea
-(0.005 \pm 0.0015)~a^2 p^2-(1.9 \pm 0.4)~GeV^3 ~a/p^2
\eea
In the present case, the role of these artefacts is crucial to obtain the condensate. In fact, 
before extraction of these artefacts, the clover data show a rather flat behavior at large $p$, or even an increase at $6.8$.
The term with $a/p^2$ considerably improves the fit, with a $\chi^2$ divided by 3, and brings in a value of the condensate
quite close to the one obtained with overlap fermions .
These results seem a signal that we have obtained a rather
accurate treatment of artefacts. Of course, one may be worried of having introduced a non canonical artefact, which was not necessary in the overlap
case ; yet, we must remember that there is no logical reason why such terms should be absent (in the overlap case, they are present anyway in $g_1$).

With a subleading term 
\bea
\propto 1/p^4
\eea
we obtain a still somewhat better agreement with the overlap condensate :
\bea
\langle A^2 \rangle=(2.83  \pm   0.35)~GeV^2 
\eea
The subleading term is also of same sign and comparable magnitude as for overlap action :
\bea
 (-1.85   \pm   0.85)~GeV^4 /p^4
\eea

Note that the consistency of the Wilson $24^4$ results with the overlap analysis is comforting the idea that the volume
effects are not too important in the overlap case, although certainly they are less compelling
because of the need to introduce the $a/p^2$ artefact.

\section{Consistency with the gluon analysis} \label{gluon}

To stress the overall consistency of the analysis, one has also to consider the
agreement with the previous gluon analysis. We quote the result of a combined fit of $\langle A^2 \rangle$ and $\Lambda_{QCD}$ to the gluon propagator and the symmetric three-gluon vertex \cite{boucaud0101302} :
\bea
\langle A^2 \rangle=(3.6 \pm 1.2)~GeV^2 ~(~\alpha_{MOM}~),~(2.4 \pm 0.6)~GeV^2 ~ (~\rm{gluon~propagator}~)....
\eea
with $\Lambda_{QCD}=0.233 \pm 0.028~GeV$, and three-loop anomalous dimensions
(the use of our present $\Lambda_{QCD}=0.237$ would lower somewhat the condensate values). The fit has been done with the same convention as for quarks, that the Wilson coefficient (calculated only to one loop) is expressed in terms of $\alpha_s^{MOM}$.
(We could get free from this convention by comparing the magnitude of power corrections themselves as in eq. \ref{cMOM}). The coincidence of the values of $\langle A^2 \rangle$ with the quark value is highly significant since it concerns the continuum
function, extracted by a series of various, independent, manipulations committed
on the gluon and quark Green functions, and since these continuum functions
are related only thanks to the OPE. 

The quark measurement has much smaller statistical errors. This is simply due to the fact that in the gluon case, we have left free the value of $\Lambda_{QCD}$, and moreover, the value of the condensate depends strongly on this value, whence the large errors in the gluon case. On the contrary, in the quark case, we have chosen a fixed $\Lambda_{QCD}=0.237~GeV$. Indeed, in this case, the dependence on $\Lambda_{QCD}$ is rather weak. Then it is useless to try to determine $\Lambda_{QCD}$ from the fit, and on the other hand the value obtained for $\langle A^2 \rangle$ remains well determined if we allow some variation in $\Lambda_{QCD}$. Another advantage of quarks is that we can reach the accuracy of four loops in the theoretical expression for the purely perturbative part. In the gluon case, such an accuracy is only possible for the asymmetric vertex, but in that case, the leading term in OPE is not given by $\langle A^2 \rangle$ \cite{desoto0105063}.

\section{Conclusions and discussions}
\label{concl}

\subsection{Physical results and evidences of various artefacts}

Let us summarize our results in the following few points :

\begin{itemize}

\item First of all, and this is the main point, let us stress that 
we have finally obtained
a rather non trivial confirmation of the validity of the OPE in the 
non gauge-invariant sector of {\it lattice} QCD (treated numerically). The virtue of OPE is that one
can describe the departure of all the various Green functions from the perturbative 
approximation at large momenta with the same set of expectation values. 
We have now obtained consistency not only between gluon Green functions, but also
with the quark sector. This is highly non trivial. Let us recall that there
is often some doubt raised about OPE itself, and the possibility is sometimes considered that
power corrections could be present not corresponding to an operator v.e.v. The final
consistency of the determination of the condensate from three distinct type of Green functions
strongly suggests that this is not the case, at least for leading power $1/p^2$
corrections. We do identify an OPE power correction, consistently related to $A^2$.
$\langle A^2 \rangle$ is found consistently large and positive. The precise magnitude of $\langle A^2 \rangle$ is affected by an important uncertainty, due to the low accuracy of the theoretical calculation of the Wilson coefficient. But the power correction (i.e. the product of the coefficient and the condensate) is well determined by the lattice analysis, and the ratio of the power corrections in the various Green functions is
actually as expected from lowest order OPE.

\item It turns out that the lattice discretisation artefacts
are unusually sizable in the quark propagator $Z_{\psi}$, but a clearcut distinction
must be made between hypercubic artefacts, which are gigantic, but can be efficiently eliminated, and the $O(4)$ symmetric ones, which are not so catastrophic, 
but that we have not been able to handle systematically.

\item  We believe that we have been really efficient in 
getting rid of the hypercubic artefacts thanks to our 
(improved) method of "restoration of $O(4)$ symmetry".

\item  Once these artefacts  have been subtracted, 
the overlap $Z_V(p^2)\equiv Z_\psi(p^2)/g_1(p^2)$, which should be independent of $p$ except for artefacts, is very close to a constant 
at large $p>5~GeV $. This is far from trivial and
supports the statement that we have no remaining $O(4)$ symmetric artefacts
in this specific region. This is directly supported by the near constancy of 
the quantities as function of $\beta$.

Moreover, also in the overlap case, but for $Z_\psi(p^2)$ only, 
the same statement of constancy with $\beta$ extends down to the lowest momenta ; this leads to suspect that $O(4)$ symmetric artefacts are small in this case
over the whole range of $p$. We are unable to explain these two special situations . 

Let us recall that a certain flatness $Z_V(p^2)$ at large $p$ is
also observed with the Wilson quark action, although it is not
so good. Let us recall also that this region is the basis for the 
standard determinations of MOM renormalisation constants, usually 
with the Wilson quark action (see for example \cite{becirevic-lat0401033}, which presumes that discretisation 
artefacts are not
large there. Let us then recall that we have no theoretical
argument supporting the statement that they are not large.
Quite the contrary. If any, the theoretical arguments
would suggest them to increase with $p$. The support
is purely empirical. This is embarrassing
if we aim at precision determinations.

\item $Z_V$ of the overlap action at $s=0$ 
is large, around $1.8$, in qualitative agreement with
BPT perturbation theory which finds large self-energy contributions
in $Z_\psi(p)$.
But it is still much larger than the expectation, and the lack of dependence 
on $\beta$ is not understood from perturbation theory.

\item Considering the cases where the $O(4)-symmetric$ artefact-free results are supected to be small, we try to fit them by OPE, i.e. by the four-loops perturbative
contribution plus the $<A^2>$ condensate contribution computed to 
leading logarithm. The overlap $Z_\psi$
and $g_1$ at large $p>5~GeV$ allow for a good fit for $\beta=6.4,6.6,6.8$,
leading to a consistent $<A^2>$ not far from $3~GeV^2$.
The overlap $Z_\psi$ 
also allow for a good fit for the whole range $p>2.6$ GeV , including in addition $\beta=6.0$.
The $<A^2>$ condensate is consistent with the former value. A very small $1/p^4$ term still improves the consistency.

\item In the other cases, namely in $g_1$ and $Z_V$ for $p$ lower than $p\sim 5~GeV$,
the $O(4)-symmetric$ artefacts become large, especially at small $p$, and in fact,
they increase regularly from large to small $p$. This trend, which is 
also contrary to the expectation of lattice perturbation theory, clearly indicates
a non perturbative origin. In fact, one important conclusion of our study is
the existence of these very large {\it non perturbative artefacts at small $p$}
due to {\it discretisation}. These are very embarrassing for any analysis
of the Green functions, as we comment below.

Let us repeat our strong conviction that these low $p$ effects(under $2~GeV$) are indeed {\it discretisation} artefacts and not volume artefacts. They behave quite counter to volume effects, as we have extensively argued.  

\item A short study of $Z_A^{MOM}/Z_V^{MOM}$ shows that similar artefacts
are still present in a ratio where they would be expected to cancel if one
applies naively the exact chiral symmetry of the lattice action. They are
clearly seen as being discretisation artefacts because they are reduced at
larger $\beta$. They seem to be present on top of an actual continuum
effect, small but visible, which could be due to continuum chiral symmetry 
spontaneous breaking. One can suspect that the artefacts themselves are connected
with the spontaneous breaking of chiral symmetry.

It seems logical
that such chiral-symmetry violating artefacts, as well as the continuum effect,
be only forbidden at large
$p$, if they are connected with spontaneous breaking effects. Indeed, it is there and only
there that 
such effects fade away. Therefore, it is
consistent with this interpretation
that we find $Z_A^{MOM}/Z_V^{MOM}=1$ at large $p$ to a high precision,
without $\beta$ dependence.

\item Of course, the clover (SW) action has the great advantage that it does not
present the same very constraining limit in volume as the overlap one ; however, it leads to less compelling results than the overlap 
one, because we have not found here the same particular situations where
$O(4)$ artefacts can be neglected ; the small $p$ artefacts, which render so difficult the determination of power corrections, are present in $Z_{\psi}$ and not only in $g_1$. As explained above, by including more and more terms to describe these artefacts, we destabilize the numerical value of $\langle A^2 \rangle$. Stopping with the first term $a/p^2$, we obtain
consistency with the overlap results.

\item As to comparison with other works on the quark propagator, 
the question of the presence and magnitude of power corrections is a crucial test of the precision obtained in the treatment of Green functions : the condensate value $\langle A^2 \rangle$ should be independent of the action and of $\beta$ with due renormalisation. In our opinion, safe and accurate extraction of power corrections requires a very large range of momenta, and therefore a large range of $\beta$ \footnote{ The lattice group in Adelaide has performed extensive studies of the
quark propagator (see references above). Recently, the same group has extended its analysis  to overlap action \cite{zhang0301018} and a special gauge action, with several $\beta$'s, corresponding to $a^{-1}\sim 1~GeV^{-1}-2~GeV^{-1}$, yet notably lower than our largest cutoffs.} ; indeed, to use a rather large range of momenta at a fixed $\beta$ would be dangerous because of the periodicity of the lattice ; the large $p$ behavior would then be highly dependent on empirical redefinitions of the momenta, aiming to remove empirically lattice artefacts at very large $p \simeq 10~GeV$.  Of course, a crucial question is whether one can work with a $\beta$ as large as $6.8$ with our lattice size $16^4$- the possible size is indeed strongly restricted for the overlap action. It is our conviction, for reasons which have been explained in detail. Another concern is that, according to our experience, to extract the real power corrections, one needs a particularly careful elimination of hypercubic artefacts.

\item The resulting value of $\langle A^2 \rangle$ from the {\it OPE} analysis of lattice QCD data should be compared to tentative estimates made by various authors within analytical approaches. One will find abundant references in the paper of Dudal et al., ref. \cite{dudal-th0311194}. It is clear that this comparison must take care of the precise definition of the condensate, as regards for instance renormalisation.

\end{itemize}

\subsection{Systematic errors on $\langle A^2 \rangle$}

Of course, we are making many assumptions which introduce uncertainty in the
value of $\langle A^2 \rangle$. Recall that we do not claim to determinate it only from the present study, since we have several previous determinations from the gluonic sector. So we can also appreciate systematic errors from the consistency we obtain with these previous estimates 
(see above, section \ref{gluon}). In fact, as we have observed in section \ref{gluon}, the gluon propagator determination and the one from $\alpha_s$ are notably different, but the values are affected by very large errors, and are compatible with the present ones. From all the determinations, we could conclude that the systematic errors do no seem to exceed $1~GeV^2$. But there is in fact an important source of systematic error, which is explained below, and which cannot be estimated by comparison with gluons because it is present in both: it is the fact that the Wilson coefficient is calculated only at low order. Then, it remains useful to discuss the sources of errors
inside the quark sector itself, for which anyway the conditions are intrinsically very favourable. 

Since we do not claim to do phenomenology, but rather an exercise in quenched QCD, we have not to bother about the quenched approximation, which is also
supposed for gluons. Chiral limit is assumed on the theoretical side, for instance to calculate the Wilson coefficents. Now, of course, we do not work at zero quark mass on the lattice. We have not tried to do a systematic chiral extrapolation on the lattice data, which would only lead to increase the statistical
errors. We observe that $Z_{\psi}$ seems very weakly dependent on our set of masses, which means that this limit is not a priori a problem at the smallest mass (at which we have made all our OPE analysis). Anyway, we can discard 
any catastrophic effect at very low quark masses through the consistency with
the quenched gluon data.

Let us now pass to more relevant effects. 

Some come from the treatment of artefacts, for
which we lack of theoretical basis, some are relative
to our description of the continuum, which, although based on a much stronger 
theoretical basis, 
involves necessarily approximations.

We do not return to finite volume artefacts, since we have nothing quantitative to say about their magnitude in the overlap case.
For what concerns hypercubic artefacts, we may have an idea on the error
remaining in their treatment by the variation observed with two variants, with a slightly different description of the continuum limit :
\bea
\langle A^2 \rangle=(3.23-3.0)~GeV^2
\eea
Whenever large $O(4)$ symmetric artefacts are present, we are
compelled, as we have seen, to rather arbitrary assumptions on their structure, 
since we have concluded that we cannot rely
on lattice perturbation theory. The correlated errors
in our determination of the "non perturbative" artefacts 
and of the condensate seem very
large, as judged from the range of values obtained in various fits, 
to such a point that we have renounced to extract any number in this case. Therefore,
we consider only the case where we have strong hints that the artefacts are small,
in which case we have mainly to consider uncertainties in the {\it continuum} description. 
They are themselves of two origins : the perturbative calculation of 
Wilson coefficients ; the non perturbative aspect, i.e. the relevance of the OPE expansion, the enumeration of operators... 
As to uncertainties in perturbative calculations :

i) We have checked that computing the perturbative contribution to third or fourth 
order in perturbation does not
change significantly the estimated condensate (only $7 \%$ of change). Another test is to reexpress the series in terms of $\alpha_{\overline{\rm MS}}$ instead of $\alpha_{\rm MOM}$. $Z_{\psi}$ changes by less than $1\%$ with various prescriptions. We can thus assume that
the perturbative contribution has been expanded far enough.

ii) As we have explained in subsection \ref{power}, the problem is much more important for the Wilson coefficient of the $A^2$ operator which has, on the contrary,
only been computed to leading logarithm.  A sign of this problem is seen by changing $\alpha_{\rm MOM}$ into
$\alpha_{\overline{\rm MS}}$. A change of $\alpha_{\rm MOM}$ into
$\alpha_{\overline{\rm MS}}$ reduces it by fourty percent $\alpha_s(10 {\rm GeV})$, and more for the smaller momenta; whence a reduction of the Wilson coefficient by $50\%$ in average. Through a conspiracy with the smaller change in the perturbative contribution, this change amounts to an
increase of the resulting condensate by $70 \%$. Of course, similar effects are present for gluons, and, as we have explained, the ratio of condensates obtained from quarks and gluons will remain the same.
More importantly, one must be aware that the {\it power correction by itself remains  well
determined}; what is not well determined is the translation of the power correction into a $\langle A^2 \rangle$ condensate value. Indeed, this translation depends on the theoretical
evaluation of the Wilson coefficient, which is not accurate at present.  

As to the properly non perturbative aspect, we may think of two sources of uncertainties in determining $\langle A^2 \rangle$.
The one stems from other operators which could enter with the same power in the OPE expansion.
However, we do not find any such operator contribution in $Z_{\psi}$ in the chiral limit. We could have some contamination since we are not exactly in the chiral limit, but it must be small, since we observe a very weak quark mass dependence. The other could be the possibility that $p$ is not sufficiently large
for the leading correction $1/p^2$ to completely dominate over next ones. This possibility is represented by the $1/p^4$
term and the fact that it is small but non zero in the fits shows indeed 
that we are not completely asymptotic at such large momenta ; although it is very small, around $1\%$ at $3~GeV$, it leads to a change of around $15 \%$ in $\langle A^2 \rangle$. On the other
hand, with this term included, we find a very good stability of $\langle A^2 \rangle$ over the large range
$2.6~GeV<p<10~GeV$, when varying the fitting window, which suggests that we have
correctly accounted for the small subasymptotic effects. Actually, this term could also mimick a neglected logarithmic dependence ; indeed, passing from $\alpha^{MOM}$ to $\alpha^{\overline {MS}}$ as expansion parameter as explained above, the $1/p^4$
term passes from $-1.1/p^4~GeV^4$ to $-(0.65 \pm 0.34)~GeV^4/p^4$, therefore there is an
appreciable variation, although the sign and order of magnitude are encouragingly stable.

On the whole, the dependence of the Wilson coefficient on the scheme for $\alpha_s$ seems the most worrying source of uncertainty, yet it can be solved soon. 

Another concern is the value used for $\Lambda_{QCD}$. Let us vary by $\pm 10\%$
our $\Lambda_{\overline {MS}}=0.237~GeV$. We find :
\bea
\Lambda_{\overline {MS}}=0.215-0.260~GeV~\to~
\langle A^2 \rangle=3.45-3.02~GeV^2,
\eea
a quite moderate change indeed. There is naturally an increase for decreasing $\Lambda_{QCD}$ because the larger power correction compensates for the slower falloff
of the perturbative part.

\subsection{General consequences for lattice studies induced 
by the observed discretisation artefacts and power corrections. Accuracy on Green functions. Renormalisation
in the MOM approach} 

True, the direct object of this study has been  
to verify the consistency of our OPE analysis of lattice data by extending it to the quark sector, and then comparing with the previous analysis of the gluon sector; 
and thereby to assess the
soundness of our statement of large power corrections
in "elementary" Green functions
as well as of their interpretation in terms of the
non gauge-invariant condensate $<A^2>$.

However, one must be aware of the strong consequences of this study,
as well as of the preceding ones, on general problems,
especially in precision studies : 

-1) {\it Power corrections.} First, we have a problem independent of the discretisation of the action : the presence
of the power corrections has the effect of modifying the estimate
of quantities defined in the perturbative
regime of QCD, when one attempts to extract them from the numerical measures done on the lattice. Indeed, such power corrections, of non perturbative origin, must
be necessarily subtracted from the Green functions to get the
perturbative contribution, the only one which presents a universal character
since it can be translated from one renormalisation scheme to another.

In the gluonic sector, the perturbative contribution leads to a determination of $\Lambda_{QCD}$ ; the necessary extraction of the $1/p^2$ power correction induces a striking modification in the
value of $\Lambda_{QCD}$, as was found some years ago \cite{boucaud0003020}. 
Of course, this very large change of $\Lambda_{QCD}$ corresponds to
much more moderate corrections on the Green functions themselves,
yet they may amount to several percents ($10\%$ at low points).

Now, let us recall that for {\it renormalisation constants} of the quark sector, 
one is also most often looking for the ones defined in the {\it perturbative regime}.
Indeed, only such perturbative renormalisation schemes can be connected
between one another by analytical calculations, and also connected
to Wilson coefficients in order to produce physical quantities.
But we predict from OPE, and we have indeed observed in this article, that power corrections of the same magnitude
affect the quark sector, in particular $Z_{\psi}$ and various vertex functions. From the Ward identity, it happens that for quantities
like $Z_{V}$, the $A^2$ power corrections cancel between the vertex
function and $Z_{\psi}$. But this is not true in general : in $Z_S$ and $Z_P$ for
instance, this cancellation does not occur. In such cases, the power corrections must be subtracted,
in principle. Of course, one may wonder whether this is practically
important. It depends on the accuracy we want to obtain. If we aim
at a precision of a few percent, certainly we do require to take them into account, since they reach several percent around $4~GeV$, $5\%$ on $Z_{\psi}$
in the Wilson case. Now, one often
claims to go below $10\,\%$ with dynamical quarks ; then,such effects are deserving of consideration. 

What simplifies somewhat the problem raised by power corrections is that not only they are independent of the chosen discretisation of the action, but 
they are often related to the same $\langle A^2 \rangle$  condensate, at least as regards the dominant power. Once $\langle A^2 \rangle$ has been confidently determined by one analysis, it can be used in others, the respective contribution to the various Green functions being obtained through the lowest order Wilson coefficients of the OPE.

-2) the $O(4)$-invariant {\it discretisation} artefacts that we have found set a more general difficulty, and one which is more embarrassing, in view of our lack of theoretical control.
We have no reason to suppose that the special case we have studied is especially
catastrophic. Yet, it shows already an embarrassing situation.

$\alpha$-The first step is to have control on {\it hypercubic} artefacts
to a good accuracy.
It is deserving of mention that this accuracy cannot be obtained by the standard
method of selecting democratic points. Moreover, the simplest versions of our alternative method of "restoration of $O(4)$ symmetry" have not allowed to get a good accuracy. We have had to go further. This requires already a good deal of work. But, finally, it seems that we are able to produce a {\it systematic and accurate} procedure.

$\beta$-On the other hand, we have not obtained a safe a systematic general method to extract the $O(4)$-invariant artefacts while we have a clear proof that they are large at small $p$. This is disastrous for the extraction of the power corrections.
This fact seems to be a problem specific to the {\it quark} sector : we had not found
similar evidence for gluonic Green functions, although we cannot exclude totally
their presence. 
We have not any theoretical mastering about their magnitude. It is our impression that the perturbation theory is of no help,
since they happen in fact to be large at small $p$, and small at large $p$,
counter to the expectation of  perturbation theory. There is no explanation for
the relative flatness of $Z_{V}$ we observe at large $p$, and which is the basis of most determinations of renormalisation constants. Note that this lack of explanation is a
problem not only for us, but for all the $MOM$ determinations of renormalisation constants.

One may wonder whether the choice of the quark {\it action} may help.
The use of {\it overlap} action seems to introduce larger hypercubic artefacts, but this is not a decisive obstacle, as we have explained. On the other hand,
the fact that in certain cases it presents small $O(4)$-invariant artefacts seems an important advantage. Yet, it is weakened by our lack of understanding of the underlying reasons why it is so. Calculations with other values of $\rho=1+s$ may reveal instructive. 

Taking into account these uncertainties
coming from $O(4)$-invariant artefacts, 
which  have appeared negligible only for $Z_{\psi}$ and only in the overlap case, but not at all for $Z_V$, it is possible that the MOM scheme may reveal not very practicable
for precision calculations, although it is appealing by its simplicity, and quite efficient for ordinary purposes. Of course, at this point, one may think of the method of the ALPHA collaboration as a complementary one, technically difficult, but which allows a very clean treatment of discretisation artefacts by using on-shell quantities, and also allows to work at very high energy scale,
therefore eliminating power corrections (see for example, for $Z_V$, \cite{luscher9611015}).

\section{Acknowledgement}

We thank Michele Pepe, Damir Becirevic, Claude Roiesnel.
for precious discussions and comments, and M. Pepe for help when initiating this work. Alain Le Yaouanc would like to thank J.-R. Cudell and C. Pittori for many very useful discussions in a previous collaboration on the quark propagator. 
This work has been supported in part by the European Network 
``Hadron Phenomenology from Lattice QCD'', HPRN-CT-2000-00145.
 We have used for this work 
the APE1000 hosted by the Centre de Ressources
Informatiques (Paris-Sud, Orsay) and purchased thanks to a funding from the
Minist\`ere de l'Education Nationale and the CNRS.

\appendix

\section{Perturbative expansion to four loops in the full MOM scheme}
\label{A}


Our aim is to express $Z_{\psi}(p^2)$, in terms of 
$\alpha_s^{MOM}$, defined by the triple gluon vertex at symmetric momenta. More precisely, we are looking for the renormalisation group 
improved expression, which resums the large logs $\log (p^2/\mu^2)$
at large $p$. The expression then takes the form of a series in $\alpha_s^{MOM}(p)$ with pure number coefficients. This series can be obtained from the knowledge of the anomalous dimension of $Z_{\psi}$ in terms of $\alpha_s^{MOM}$, and of the $\beta$ function of the MOM scheme. This is possible in the Landau gauge to the order of four loops included, thanks to the papers of Chetyrkin and collaborators, ref. \cite{Chetyrkin:hep-ph/9910332} and \cite{Chetyrkin:hep-ph/0008094}. 

First, one has in section 4.2 of the first paper the 
expansion of $\frac {\partial ln Z_{\psi} (\mu^2)} {\partial ln \mu^2} $ as a
series in $\alpha_s$ of the $\overline {\rm {MS}}$ scheme to four loops (in fact the authors consider the inverse $Z_2(\mu^2)=(Z_{\psi} (\mu^2))^{-1}$, so we have to invert their formula). Since there is no $\alpha_s$ term in the Landau gauge, to reexpress the series in terms of the $\alpha_s$ of the ${\rm MOM}$ scheme at four loops requires the expansion of $\alpha_s^{\overline {\rm MS}}$ in terms of $\alpha_s^{\rm MOM}$ only to order three included. This expansion is provided by the section 5 of the second paper, by inverting the first equation of this section. On the other hand, we need the MOM $\beta$ function. It is given in the second paper, section 6, at three loops, which is also sufficient to calculate the renormalisation improved series for $Z_{\psi}(p^2)$ at four loops included ; indeed, the four loop coefficient of the $\beta$ function enters only as a factor of the one loop anomalous dimension of the fermion $\gamma_0$ , which is zero in the Landau gauge. We then obtain the following expansions :

-$Z_{\psi}$ is expressed in general as :
\bea\label{zpsiI} 	
	Z_\psi^{pert} &=& \alpha^{\frac {\gamma_0} { \beta_0 }}\Bigg( 1 
           + \alpha  \frac{ \beta_0  \gamma_1 - 2  \beta_1  \gamma_0 } 
           { 4  \pi  \beta_0^2 }\quad\quad\quad\quad\quad \nonumber \\
           &+& 0.5  \alpha^2  \left( \frac{( \beta_0  \gamma_1 
           - 2  \beta_1  \gamma_0 )^2 }{ 16  \beta_0^4  \pi^2 }
           + \frac{ 8  \beta_1^2  \gamma_0 - \beta_0  \beta_2  \gamma_0  
           - 4  \beta_0  \beta_1  \gamma_1 + 2  \beta_0^2  \gamma_2 }
             { 32  \beta_0^3  \pi^2 } \right) \nonumber \\
          &+& \frac{ \alpha^3} { 768  \beta_0^6  \pi^3 }   \Big(
           - 16  \beta_1^3  \gamma_0^3 + 24  \beta_0  \beta_1^2 
            \gamma_0^2  ( - 2  \beta_1 + \gamma_1 ) \nonumber \\
	    &+& 2  \beta_0^2
            \beta_1  \gamma_0  ( - 16  \beta_1^2 + 3  \beta_2  \gamma_0
           + 24  \beta_1  \gamma_1 - 6  \gamma_1^2 ) \nonumber\\
           &-& 2  \beta_0^4  ( 2  \beta_3  \gamma_0 + \beta_2  \gamma_1
           + 4  \beta_1  \gamma_2 - 3  \gamma_1  \gamma_2 ) \nonumber \\&+& \beta_0^3
            \left( 16  \beta_1^2  \gamma_1 - 3  \beta_2  \gamma_0  \gamma_1
           + 2  \gamma_1^3 + 4  \beta_1  \left( 2  \beta_2  \gamma_0 
          - 3  ( \gamma_1^2 + \gamma_0  \gamma_2 ) \right) \right)  
           + 4  \beta_0^5  \gamma_3 \Big)\Bigg)
\eea     

-For $n_{\rm f}=0$  in the Landau gauge, we get :   
\bea
	\gamma_0 &=& 0.0\quad
	\gamma_1 = \frac{67}{3}\quad
	\gamma_2 = -94.7943 \nonumber\\
\gamma_3 &=& 14503.7
\eea

-$\alpha$ , which represents here $\alpha_{{\rm MOM}}(p)$ (defined through the three-gluon vertex at symmetric momenta), is given, under the same conditions, by :
\bea	
	\alpha(p) &=&  \frac{4 \pi}{\beta_0 \, t} 
     	  -  8  \pi \frac {\beta_1} {\beta_0}  \frac{\ln t} {(\beta_0  t )^2} 
     	  + \frac{1} { ( \beta_0  t )^3}  \left( 
     	  2  \pi \frac{\beta_2}{\beta_0} + 16  \pi  \frac{\beta_1^2} 
     	  {\beta_0^2}  \left( (\ln t)^2 - \ln t - 1 \right) \right)
     	    \\
	  &+&\frac{1} {( \beta_0  t )^4} \left(
      	  - 32  \pi  \frac{\beta_1^3} {\beta_0^3} (\ln t)^3
     	  + 80  \pi \frac{ \beta_1^3}{\beta_0^3} (\ln t)^2 \right. \nonumber \\
     	  &-& \left. \frac{ 12 \pi  \beta_0  \beta_1  \beta_2 - 64  \pi  \beta_1^3}
	{  \beta_0^3} \ln t
     	  + \frac{ 2  \pi  \beta_0^2  \beta_3 - 16  \pi  \beta_1^3 }
     	  { \beta_0^3} \right) \nonumber 
  \eea   
where $t= \ln (p^2/\Lambda_{{\rm MOM}}^2)$ (in the MOM case, one stops
at the $1/t^3$ terms.
and where
\bea   
    \beta_0 = 11  \quad
\beta_1 = 51 \quad 
\beta_2 = 3072~(10\%)
\eea


\section{$A^2$ Wilson coefficient in the quark propagator}      
\label{B}

In order to renormalise the bare quark propagator in Eq. (\ref{barecondensat}) we will define the
following two renormalisation constants, both in the momentum subtraction (MOM) general scheme :

\bea
\left( \begin{array}{l}
Z_\psi(\mu^2) \\ \\ Z^{\rm pert}_\psi(\mu^2) 
\end{array} \right) \ \delta_{a,b} \ = \ \frac{-i\pslash + m(p^2)}{p^2 + m(p^2)} 
 \ \left( \begin{array}{l}
S^{-1}(p) \\ \\ S^{-1}_{\rm pert}(p) 
\end{array} \right)_{p^2 = \mu^2}
\label{Zs}
\eea

\noindent The constant in the top of the l.h.s of Eq. (\ref{Zs}) includes the non-perturbative 
contributions to the quark propagator to let it take the tree-level value all over the energy 
range (not only in the perturbative regime). 
We renormalise Eq. (\ref{barecondensat}) by multiplying by that of the bottom 
(this purely perturbative MOM renormalisation constant, computed to four loops in 
ref. \cite{Chetyrkin:hep-ph/9910332}, is presented in the appendix \ref{A}) ; as to the Wilson coefficient of $A^2$, we calculate it at leading RG order.

\bea\label{renorm}
\left( Z^{\rm pert}_\psi(\mu^2) \right)^{-1} \  S^{-1}(p) \ = \ 
i \pslash \ \delta_{a,b} \frac{Z_\psi(p^2)}{Z_\psi^{pert}(\mu^2)} + \cdots \nonumber \\
= \ \left( Z^{\rm pert}_\psi(\mu^2) \right)^{-1} \ S^{-1}_{\rm pert}(p) + 
        i \pslash \ \frac{d\left(\frac{p^2}{\mu^2},\alpha(\mu) \right)}{p^2} 
        \frac{< (A^2)_{\rm R}(\mu) >}{4 (N_C^2-1)} \ \delta_{a,b} + \cdots \nonumber \\
= \ i \pslash \ \delta_{a,b} \ \left( \frac{Z_\psi^{pert}(p^2)}{Z_\psi^{pert}(\mu^2)} + 
        \frac{d\left(\frac{p^2}{\mu^2},\alpha(\mu) \right)}{p^2} 
        \frac{< (A^2)_{\rm R}(\mu) >}{4 (N_C^2-1)}
         \right) \ + \cdots
\eea
where $(A^2)_{\rm R}(\mu^2) = Z_{A^2}^{-1}(\mu^2) : A^2_{\rm bare} :$ \footnote{the : $\cdots$  fixes 
the normal order substracting the additive divergencies in $A^2_{bare}$ (see discussion in ref. 
\cite{A2bare})}, and where we only write the leading terms in $\pslash$. 
Concerning $d\left( \frac{p^2}{\mu^2},\alpha(\mu) \right)$, the same procedure used for gluon Green 
functions in refs.\cite{boucaud0008043,boucaud0101302,desoto0105063} is in order here. Then, from 
Eq. (\ref{renorm}), multiplying by $Z^{\rm pert}_\psi(\mu^2)$, and taking logarithm derivatives on $\mu$ in both sides, we obtain the following RG equation:
\bea
\left\{
- \gamma_0 \frac{\alpha(\mu)}{4\pi} + \gamma_{A^2}^{(0)} \frac{\alpha(\mu)}{4\pi} 
+ \frac{d}{d \ln{\mu^2}}  \right\} d \left( \frac{p^2}{\mu^2},\alpha(\mu) \right) \ = \ 0 
\eea


\noindent that is identically satisfied with :

\bea\label{ansatz}
d\left( \frac{p^2}{\mu^2},\alpha(\mu) \right) \ = \ d\left( 1,\alpha(p) \right) \
\left( \frac{\alpha(\mu)}{\alpha(p)}\right)^{\frac {-\gamma_F^{(0)} +\gamma_{A^2}^{(0)}} {\beta_0} } \ ,
\eea
\noindent where $d(\left( 1,\alpha(p) \right))$ is in fact beginning with $\alpha(p)$ and $\gamma_{A^2}^{(0)}$ is defined through :

\bea
\frac{d}{d\ln{\mu^2}} \ln{Z_{A^2}} \ = \ - \gamma_{A^2}^{(0)} \frac{\alpha(\mu)}{4\pi} + {\cal O}\left(\alpha^2\right) \ ;
\eea

\noindent and where all the involved one-loop coefficients are well known:

\bea\label{diagram}
\gamma_{A^2}^{(0)}= \frac{35 N_c}{12} \ , \ \ \
\gamma_0=0 \ , \ \ \
\beta_0=11 \ .
\eea
On the other hand, $d\left( 1,\alpha(p) \right)$ can be obtained by computing 
the only diagram involved in the ``OPE business'' for this case,

\vspace*{1.25cm}
\begin{center}
\hspace*{-4cm} \SetScale{0.6} \quark 
\end{center}

\vspace*{-3.5cm}

\bea
\ \ \ \ \ \ \ \ \ \ \ \ \ \ \ \ \ \ \ \ \ \ \ \
\to  \ \ \ \ \ d\left( 1,\alpha(p) \right) = \frac{(N_C^2-1)}{3} g^2(p)
\eea

\vspace*{0.8cm}

\noindent Then, by applying the results from Eqs. (\ref{ansatz}-\ref{diagram}) 
to Eq. (\ref{renorm}), we will obtain the final result Eq. (\ref{condensat}) to be 
used for the fits.


\end{document}